\documentclass[preprint,pre,aps,showpacs]{revtex4}
\usepackage{graphicx}
\usepackage{dcolumn}
\usepackage{bm}
\usepackage{amsmath}
\usepackage{color}
\begin{document}
\title{Equilibration and aging of
liquids of non-spherically interacting  particles}
\author{Ernesto C. Cort\'es Morales$^{1,a}$, L.F. Elizondo Aguilera$^{2,3,b}$,
and M. Medina Noyola$^{1,2}$}

\affiliation{ $^{1}$ Instituto de F\'isica Manuel Sandoval Vallarta,
Universidad Aut\'onoma de San Luis Potos\'i, Alvaro Obreg\'on 64,
78000 San Luis Potos\'i, SLP, M\'exico.}
\email{a) ernie2k@ifisica.uaslp.mx}
\affiliation{  $^{2}$ Departamento de Ingenier\'ia F\'isica, Divisi\'on
de Ciencias e Ingenier\'ias, Universidad de Guanajuato, Loma del Bosque 103,
37150 Le\'on, M\'exico.}
\affiliation{$^{3}$ Institut f\"ur Materialphysik im Weltraum, Deutsches Zentrum
f\"ur Luft-und Raumfahrt (DLR), 51170 K\"oln, Germany}
\email{b) luisfer.elizondo@gmail.com}
\date{\today}

\begin{abstract}

The non-equilibrium self-consistent generalized Langevin equation
theory of irreversible processes in liquids is extended to describe
the positional and orientational thermal fluctuations of the
instantaneous local concentration profile
$n(\mathbf{r},\bm{\Omega},t)$ of a suddenly-quenched colloidal liquid
of particles interacting through non spherically-symmetric pairwise
interactions, whose mean value
$\overline{n}(\mathbf{r},\bm{\Omega},t)$ is constrained to remain
uniform and isotropic, $\overline{n}(\mathbf{r},\bm{\Omega},t)=\overline{n}(t)$.
Such self-consistent theory is cast in terms of the time-evolution
equation of the covariance
$\sigma(t)=\overline{\delta n_{lm}(\mathbf{k};t)
\delta n^{\dagger}_{lm}(\mathbf{k};t)}$ of the fluctuations $\delta
n_{lm}(\mathbf{k};t)=n_{lm}(\mathbf{k};t)
-\overline{n_{lm}}(\mathbf{k};t)$ of the spherical harmonics
projections $n_{lm}(\mathbf{k};t)$ of the Fourier transform of
$n(\mathbf{r},\bm{\Omega},t)$. The
resulting theory describes the non-equilibrium evolution after a
sudden temperature quench of both, the static structure
factor projections $S_{lm}(k,t)$ and the two-time correlation function
$F_{lm}(k,\tau;t)\equiv\overline{\delta n_{lm}(\mathbf{k},t)\delta
n_{lm}(\mathbf{k},t+\tau)}$, where $\tau$ is the correlation \emph{delay}
time and $t$ is the \emph{evolution} or \emph{waiting} time after the
quench. As a concrete and illustrative application
we use the resulting self-consistent equations to
describe the irreversible processes of equilibration or aging
of the orientational degrees of freedom of a system of strongly
interacting classical dipoles with quenched positional disorder.

\end{abstract}
\pacs{23.23.+x, 56.65.Dy}
\maketitle

\section{Introduction}

The fundamental description of dynamically arrested states of matter
is a crucial step towards understanding the properties of very common
amorphous solid materials such as glasses and gels
\cite{angellreview1,ngaireview1,sciortinotartaglia}, and of more
technologically specialized materials, such as spin glasses
\cite{spinglassbook,gingras,biltmo}. The main fundamental challenge
posed by these materials derives from their inability to reach thermodynamic
equilibrium within experimental times, and from the fact that their properties
depend on the protocol of preparation, in obvious contrast with
materials that have genuinely attained thermodynamic equilibrium. Understanding
the origin of this behaviour falls outside the realm of classical and
statistical thermodynamics, and must unavoidably be addressed from
the perspective of a non-equilibrium theory
\cite{degrootmazur,keizer,casasvazquez0}. In fact, a major
challenge for statistical physics is to  develop a microscopic
theory able to predict the properties of glasses and gels in terms
not only of the intermolecular forces and applied external fields,
but also in terms of the protocol of preparation of the material.

\emph{First-principles} theoretical frameworks exist, leading to
\emph{quantitative} predictions of the dynamic properties of structural glass
forming liquids near their dynamical arrest transitions, one of the best-known
being mode coupling theory (MCT) \cite{goetze1,goetze2}. However, this theory,
as well as the equilibrium version of the self-consistent generalized Langevin
equation (SCGLE) theory of dynamical arrest \cite{todos1,todos2}, are meant
to describe the dynamics of \emph{fully equilibrated} liquids.
Hence, the phenomenology of the transient time-dependent processes, such as aging,
occurring during the amorphous solidification of structural glass formers, falls
completely out of the scope of these \emph{equilibrium} theories. Thus, it is
important to attempt their extension to describe these non-stationary non-equilibrium
structural relaxation processes, which in the end constitute the most fundamental kinetic
fingerprint of glassy behavior.

In an attempt to face this challenge,  in 2000 Latz \cite{latz} proposed a formal
non-equilibrium extension of MCT which, however, has not yet found a specific
quantitative application. In the meanwhile, the SCGLE theory has recently been
extended to describe non-stationary non-equilibrium processes in glass-forming
liquids \cite{nescgle1,nescgle2}. The resulting non-equilibrium theory, referred
to as the \emph{non-equilibrium} self-consistent generalized Langevin equation
(NE-SCGLE) theory, was derived within the fundamental framework provided by a
non-stationary extension \cite{nescgle1} of Onsager's theory of linear irreversible
thermodynamics  \cite{onsager1, onsager2} and of time-dependent thermal fluctuations
\cite{onsagermachlup1,onsagermachlup2}, with an adequate extension \cite{delrio,faraday}
to allow for the description of memory effects.

The NE-SCGLE theory thus derived, aimed at describing non-equilibrium
relaxation phenomena in general  \cite{nescgle1}, leads in particular
\cite{nescgle2} to a simple and intuitive but generic description of the
essential behavior of the non-stationary and non-equilibrium structural
relaxation of glass-forming liquids near and beyond its dynamical arrest
transition. This was explained in detail in Ref. \cite{nescgle3} in the context
of a model liquid of soft-sphere particles. The recent comparison \cite{nescgle7}
of the predicted scenario with systematic simulation experiments of the equilibration
and aging of dense hard-sphere liquids, indicates that the accuracy of these
predictions go far beyond the purely qualitative level, thus demonstrating
that the NE-SCGLE theory is a successful pioneering first-principles statistical
mechanical approach to the description of these fully non-equilibrium phenomena.

As an additional confirmation, let us mention that for model liquids with
hard-sphere plus attractive interactions, the NE-SCGLE theory predicts a still
richer and more complex scenario, involving the formation of gels and porous glasses
by arrested spinodal decomposition \cite{nescgle5,nescgle6}.  As we know, quenching
a liquid from supercritical temperatures to a state point inside
its gas-liquid spinodal region, normally leads to the full phase separation through a
process that starts with the amplification of spatial density fluctuations of certain
specific wave-lengths \cite{cahnhilliard,cook,furukawa}. Under some conditions, however,
this process may be interrupted when the denser phase solidifies as an amorphous
sponge-like non-equilibrium bicontinuous structure \cite{luetalnature,sanz,gibaud,foffi,helgeson},
typical of physical gels \cite{zaccarellireviewgels}. This process is referred to
as \emph{arrested} spinodal decomposition, and has been observed in many colloidal
systems, including colloid-polymer mixtures \cite{luetalnature}, mixtures of
equally-sized oppositely-charged colloids \cite{sanz}, lysozyme protein solutions
\cite{gibaud}, mono- and bi-component suspensions of colloids with DNA-mediated
attractions \cite{foffi}, and thermosensitive nanoemulsions \cite{helgeson}. From
the theoretical side, it was not clear how to extend the classical theory of spinodal
decomposition \cite{cahnhilliard, cook, furukawa} to include the possibility of dynamic
arrest, or how to incorporate the  characteristic non-stationarity of spinodal decomposition,
in existing theories of glassy behavior \cite{berthierreview}. In Refs. \cite{nescgle5} and
\cite{nescgle6} it has been shown that the NE-SCGLE theory provides precisely this missing
unifying theoretical framework.

Recently the NE-SCGLE theory was extended to multi-component systems \cite{nescgle4}, thus opening the
route to the description of more complex non-equilibrium amorphous states of matter. Until now, however,
the NE-SCGLE theory faces the limitation of referring only to liquids of particles
with radially symmetric pairwise interparticle forces, thus excluding its direct comparison with the
results of important real and simulated experiments involving intrinsically non-spherical particles
\cite{arrozchino, weeks} and, in general, particles with non-radially symmetric interactions.
The present work
constitutes a first step in the direction of extending the NE-SCGLE theory to describe the irreversible
evolution of the static and dynamic properties of a Brownian liquid constituted by particles with non-radially
symmetric interactions, in which the orientational degrees of freedom are essential.

More concretely, the main
purpose of the present paper is to describe the theoretical derivation of the   NE-SCGLE time-evolution equations
for the spherical harmonics projections $S_{lm,lm}(k;t)$, $F_{lm,lm}(k,\tau; t)$, and $F_{lm,lm}^S(k,\tau; t)$,
of the non-equilibrium and non-stationary static structure factor $S(\mathbf{k},\bm{\Omega};t)$   and of the
collective and self intermediate scattering functions $F(\mathbf{k},\bm{\Omega},\bm{\Omega}',\tau; t)$ and
$F^S(\mathbf{k},\bm{\Omega},\bm{\Omega'},\tau; t)$. For this, we start from the same general and fundamental
framework provided by the non-stationary extension
of Onsager's theory, developed in Ref. \cite{nescgle1} to discuss the spherical case. The result of the present
application are Eqs. (\ref{relsigmadifnonspher13})-(\ref{flmsqframenonstats}) below, in which $\tau$ is the delay
time, and $t$ is the evolution (or ``waiting'') time after the occurrence of the instantaneous temperature quench.
The solution of these equations describe the non-equilibrium (translational and rotational) diffusive processes
occurring in a colloidal dispersion after an instantaneous temperature quench, with the most interesting prediction
being the aging processes that occur when full equilibration is prevented by conditions of dynamic arrest.

Although this paper only focuses on the theoretical derivation 
of the NE-SCGLE equations, as an
illustration of the possible concrete  applications of
the extended non-equilibrium theory, here we also
solve the resulting equations for one particular system
and condition. We refer to a liquid of dipolar hard-spheres
(DHS) with fixed positions and subjected to a sudden
temperature quench. This is a simple model of the irreversible
evolution of the collective orientational degrees of freedom of
a system of strongly interacting magnetic dipoles with fixed but
random positions. Although this particular application by itself has
its own intrinsic relevance in the context of disordered magnetic
materials, the main reason to choose it as the illustrative example
is that  Eqs. (\ref{relsigmadifnonspher13})-(\ref{flmsqframenonstats})
describe coupled translational and rotational dynamics, whose particular
case $l=0$ coincide with the radially-symmetric case, already discussed
in detail in Refs. \cite{nescgle3,nescgle4,nescgle5}. Thus, the most novel
features are to be expected in the non-equilibrium rotational dynamics
illustrated in this exercise.

Just like in the case of liquids formed by spherical particles, the
development of the NE-SCGLE theory for  liquids of non-spherical
particles requires the previous development of the \emph{equilibrium}
version of the corresponding SCGLE theory. Such an equilibrium SCGLE
theory for non-spherical particles, however, was previously developed
by Elizondo-Aguilera et al. \cite{gory1}, following to a large extent
the work of Schilling and collaborators \cite{schilling1,schilling2,schilling3}
on the extension of mode coupling theory for this class of systems. Thus,
we start our discussions  in section \ref{section2} with a brief review of
the main elements of the non spherical \emph{equlibrium} SCGLE theory and
its application to dynamical arrest in systems formed by colloidal interacting
particles with non spherical potentials.

In section \ref{nonequilib} we
outline the conceptual basis and the main steps involved in the
derivation of the non-equilibrium extension of the SCGLE theory for
glass-forming liquids of non-spherical particles. In the same section we summarize
the resulting set of self consistent equations which constitutes this
extended theory. In section {\ref{agingspinglass}, we introduce a
simplified model for interacting dipoles randomly distributed in space
and apply our equations to investigate the slow orientational dynamics
as well as the aging and equilibration processes of the system near its
``spin glass''-like transitions. Finally in section \ref{conclusions} we
summarize our main conclusions.

\section{Equilibrium SCGLE theory of Brownian liquids of non-spherical particles.}\label{section2}

In this section we briefly describe the equilibrium SCGLE theory of the dynamics of liquids formed
by non-spherical particles developed by Elizondo-Aguilera et al. \cite{gory1}.  We first describe
the main properties involved in this description and then summarize their time-evolution equations,
which constitute the essence of the SCGLE theory.

\subsection{Collective description of the translational and orientational degrees of freedom.}\label{subsection2.1}

Let us start by considering a liquid formed by $N$ identical non-spherical colloidal particles in a volume $V$
\cite{gory1}, each having mass $m$ and inertia tensor $\mathbf{I}$. The translational degrees
of freedom are described by the  vectors
$\mathbf{r}^N\equiv(\mathbf{r}_1,...,\mathbf{r}_N)$ and
$\mathbf{p}^N \equiv (\mathbf{p}_1,...,\mathbf{p}_N)$, where
$\mathbf{r}_n$ denotes the center-of-mass position vector of the
$n$th-particle and $\mathbf{p}_n\equiv
md\mathbf{r}_n/dt=m\mathbf{v}_n(t)$ is the associated linear
momentum. Similarly, the orientational degrees of freedom are described
by the abstract vectors $\mathbf{\Omega}^N\equiv
(\mathbf{\Omega}_1,...,\mathbf{\Omega}_N)$ and $\mathbf{L}^N
\equiv (\mathbf{L}_1,...,\mathbf{L}_N)$, where $\mathbf{\Omega}_n$
denotes the Euler angles which specify the orientation of the
$n$th molecule, and
$\mathbf{L}_n=\mathbf{I}(\mathbf{\Omega}_n)\bm{\omega}_n$ is the
corresponding angular momentum, so that
$\bm{\omega}_n$ denotes the angular velocity.
Let us now assume that the potential energy $U(\mathbf{r}^N,\bm{\Omega}^N)$ of the interparticle
interactions is pairwise additivity, i.e., that
\begin{eqnarray}
U(\mathbf{r}^N,\bm{\Omega}^N)=  \sum_{n,n'=1}^N u(\mathbf{r}_n,\mathbf{r}_{n'};\bm{\Omega}_n,\bm{\Omega}_{n'}), \label{totpotenergy}
\end{eqnarray}
where $u(\mathbf{r}_n,\mathbf{r}_{n'};\bm{\Omega}_n,\bm{\Omega}_{n'})$ is the interaction
potential between particles $n$ and $n'$. In the particular case of axially-symmetric
particles, that we shall have in mind here, the third Euler angle is actually redundant,
and hence, $\bm{\Omega}_n=\bm{\Omega}_n(\theta_n,\phi_n)$.

The most basic observable in terms of which we want to describe the dynamical
properties of a non-spherical colloidal system is the time
dependent microscopic one-particle density
\begin{equation}
n(\mathbf{r},\bm{\Omega};t)\equiv(1/\sqrt{N})\sum_{n=1}^N\delta
(\mathbf{r}-\mathbf{r}_n(t))\delta(\bm{\Omega}-\bm{\Omega}_n(t)).\label{localcon}
\end{equation}
Given that $\bm{\Omega}=\bm{\Omega}(\theta,\phi)$, \textit{any} function
$f(\mathbf{r},\bm{\Omega})$ can be expanded with respect to plane
waves and spherical harmonics as
\begin{equation}
f(\mathbf{r},\bm{\Omega})=\frac{1}{V}\frac{1}{\sqrt{4\pi}}\int
d\mathbf{k}\sum_{lm}(i)^lf_{lm}(\mathbf{k})e^{-i\mathbf{k}\cdot\mathbf{r}}Y^*_{lm}(\bm{\Omega}) \label{fromega}
\end{equation}
where
\begin{equation}
f_{lm}(\mathbf{k})=\sqrt{4\pi}i^l\int_{V}d\mathbf{r}\int
d\bm{\Omega}
f(\mathbf{r},\bm{\Omega})e^{-i\mathbf{k}\cdot\mathbf{r}}Y_{lm}(\bm{\Omega}). \label{flmk}
\end{equation}
Thus, using Eq. (\ref{localcon}) in (\ref{fromega}) and (\ref{flmk}),
we may define the \textit{so-called} tensorial density modes
\begin{equation}
n_{lm}(\mathbf{k},t)=\sqrt{\frac{4\pi}{N}}i^l\sum_{n=1}^N
e^{i\mathbf{k}\cdot\mathbf{r}_n(t)}Y_{lm}(\bm{\Omega}_n(t)),\label{nlm}
\end{equation}
and hence, we can define the following two-time correlation functions,

\begin{eqnarray}
F_{lm;l'm'}(\mathbf{k},\tau;t)&\equiv&\langle \delta
n^*_{lm}(\mathbf{k},t+\tau)\delta
n_{l'm'}(\mathbf{k},t)\rangle\label{flm1}\\\nonumber
&=&\frac{4\pi}{N}i^{l-l'}\sum_{n\neq n'}^{N}\Big\langle
e^{i\mathbf{k}\cdot[\mathbf{r}_n(t+\tau)-\mathbf{r}_{n'}(t)]}Y_{lm}^*(\bm{\Omega}_n(t+\tau))
Y_{l'm'}(\bm{\Omega}_{n'}(t))\Big\rangle,
\end{eqnarray}

\noindent where $\delta n_{lm}(\mathbf{k},t)\equiv
n_{lm}(\mathbf{k},t)-\langle n_{lm}(\mathbf{k},t)\rangle$. }

We also define for completeness the self components
\begin{equation}
n^S_{lm}(\mathbf{k},t)\equiv\sqrt{4\pi}i^le^{i\mathbf{k}
\cdot\mathbf{r}_T(t)}Y_{lm}(\bm{\Omega}_T(t)),
\end{equation}
and the corresponding two-time correlation functions
\begin{eqnarray}
F^S_{lm;l'm'}(\mathbf{k},\tau;t)\equiv\langle
n^{S*}_{lm}(\mathbf{k},t+\tau)n^S_{l'm'}(\mathbf{k},t)\rangle\nonumber\\
=4\pi i^{l-l'}\Big\langle
e^{i\mathbf{k}\cdot[\mathbf{r}_T(t+\tau)-\mathbf{r}_T(t)]}Y_{lm}(\bm{\Omega}_T(t+\tau))
Y_{l'm'}(\bm{\Omega}_T(t))\Big\rangle,\label{selfcorrelm}
\end{eqnarray}
where $\mathbf{r}_T(t)$ denotes the position of the center of mass
of \emph{any} of the particles at time $t$ and $\bm{\Omega}_T(t)$
describes its orientation. As indicated before, we will refer to $\tau$
as the delay (or \emph{correlation}) time, whereas for $t$ we refer to
the \emph{evolution} time.

The equal-time value of these correlation functions are
$F_{lm;l'm'}(\mathbf{k},\tau=0;t)=S_{lm;l'm'}(\mathbf{k};t)$ and
$F^S_{lm;l'm'}(\mathbf{k},\tau=0;t)=1$ where
$S_{lm;l'm'}(\mathbf{k};t)$ are the tensorial components of the
static structure factor
$S(\mathbf{k},\bm{\Omega},\bm{\Omega}';t)$. Of course, the
dependence of these quantities on the evolution time $t$ is only
relevant if the state of the system is not stationary. Under
thermodynamic equilibrium, $F_{lm;l'm'}(\mathbf{k},\tau;t)$,
$F^S_{lm;l'm'}(\mathbf{k},\tau;t)$, and
$S_{lm;l'm'}(\mathbf{k};t)$ cannot depend on $t$, and we should
denote them as $F^{(eq)}_{lm;l'm'}(\mathbf{k},\tau)$,
$F^{S(eq)}_{lm;l'm'}(\mathbf{k},\tau)$, and
$S^{(eq)}_{lm;l'm'}(\mathbf{k})$.
In Ref. \cite{gory1} the generalized Langevin equation (GLE)
formalism and the concept of contraction of the description were
employed to derive exact memory function equations for $F^{(eq)}_{lm;l'm'}(\mathbf{k},\tau)$ and
$F^{S(eq)}_{lm;l'm'}(\mathbf{k},\tau)$. These dynamic
equations only involve the corresponding projections
$S^{(eq)}_{lm;l'm'}(\mathbf{k})$ of the \emph{equilibrium} static structure
factor. For notational convenience,
however, we shall not write the label $(eq)$ in what follows,
although for the rest of this section we shall only refer to these equilibrium
properties.

As explained in Ref.  \cite{gory1}, the referred exact memory 
function equations for $F^{(eq)}_{lm;l'm'}(\mathbf{k},\tau)$ and
$F^{S(eq)}_{lm;l'm'}(\mathbf{k},\tau)$ require the independent
determination of the corresponding self and collective memory
functions. In a manner similar to the spherical case, simple
Vineyard-like approximate closure relations for these memory
functions convert the originally exact equations into a closed
self-consistent system of approximate equations for the dynamic
properties referred to above \cite{gory1}. These equations
thus constitute the extension of the \emph{equilibrium} SCGLE
theory of the dynamic properties of liquids whose particles
interact through non-spherical pair potentials.

\subsection{Summary of the \emph{equilibrium} SCGLE equations.}\label{subsection2.2}

Let us now summarize the set of
self-consistent equations that constitute the
\emph{equilibrium} SCGLE theory for a Brownian liquid of  axially-symmetric non
spherical particles. In the simplest version (we refer the reader to Ref. \cite{gory1}
for details) these equations
involve only the diagonal elements $F_{lm}(\mathbf{k},\tau)\equiv
F_{lm;lm}(\mathbf{k},\tau)$ and $F^S_{lm}(\mathbf{k},\tau)\equiv
F^S_{lm;lm}(\mathbf{k},\tau)$, and are written, in terms of the corresponding
Laplace transforms $F_{lm}(\mathbf{k},z)$ and
$F^S_{lm}(\mathbf{k},z)$, as
\begin{equation}
F_{lm}
(k,z)=\frac{S_{lm}(k)}{z+\displaystyle\frac{k^2D_T^0S^{-1}_{lm}(k)}
{1+\Delta\zeta_T^*(z)\lambda^{(lm)}_T(k)}+\frac{l(l+1)D_R^0S^{-1}_{lm}(k)}
{1+\Delta\zeta_R^*(z)\lambda^{(lm)}_R(k)}}\label{flmqframe1}
\end{equation}
and
\begin{equation}
F_{lm}^S (k,z)=\frac{1}{z+\displaystyle\frac{k^2D_T^0}{1+
\Delta\zeta_T^*(z)\lambda^{(lm)}_T(k)}+\frac{l(l+1)D_R^0}
{1+\Delta\zeta_R^*(z)\lambda^{(lm)}_R(k)}}.\label{flmsqframe2}
\end{equation}
In these equations, $D_R^0$ is the rotational free-diffusion
coefficient, and $D^0_T$ is
the center-of-mass translational free-diffusion coefficient, 
whereas the functions $\lambda^{(lm)}_T(k)$ and $\lambda^{(lm)}_R(k)$ 
are defined as $\lambda^{(lm)}_T(k)=1/[1+( k/k_{c}) ^{2}]$ and 
$\lambda^{(lm)}_R(k)=1$, where $k_c=\alpha \times k_{max}$, with $k_{max}$
being the position of the main peak of $S_{00}(k)$ and  $\alpha =1.305$. 
This ensures that for radially-symmetric interactions, we recover the 
original theory describing liquids of soft and hard spheres  \cite{nescgle4}.

On the other hand, within well
defined approximations discussed in appendix A of Ref
\cite{gory1}, the functions $\Delta\zeta_{\alpha}^*(\tau)$
($\alpha=T,R$) may be written as

\begin{eqnarray}
\Delta\zeta^*_T(\tau) =\frac{1}{3}\frac{D_T^0}{(2\pi)^3n}\int
d\mathbf{k}
k^2\sum_{l}\left[{2l+1}\right]\left[1-S^{-1}_{l0}(k)\right]^2
F^S_{l0}(k;\tau)F_{l0}(k;\tau)\label{zetart}
\end{eqnarray}

\noindent and

\begin{eqnarray}
\Delta\zeta^*_R(\tau)=\frac{1}{2}\frac{D_R^0}{(2\pi)^3}
\frac{n}{4}\frac{1}{(4\pi)^2}\int
d\mathbf{k}\sum_{l,m}\left[2l+1\right]h^2_{l0}
(k)\left[A_{l;0m}\right]^2\left[S^{-1}_{lm}(k)\right]^2
F^S_{lm}(k;\tau)F_{lm}(k;\tau)\label{zetarqq}
\end{eqnarray}
where $h_{lm}(k)$ denotes the diagonal \textit{k}-frame
projections of the total correlation function
$h(\mathbf{k},\bm{\Omega},\bm{\Omega}')$, i.e., $h_{lm}(k)$ is
related to $S_{lm}(k)$ by $S_{lm}(k)=1+(n/4\pi) h_{lm}(k)$, and
$n=N/V$ is the number density. Finally,
$A_{l;mm'}\equiv\left[C_{lm}^+\delta_{m+1,m'}+C_{lm}^-\delta_{m-1,m'}\right]$
and $C_{lm}^{\pm}\equiv\sqrt{(l\mp m)(l\pm m +1)}$.

The closed set of coupled equations in eqs.
(\ref{flmqframe1})-(\ref{zetarqq})  constitute the equilibrium non
spherical version of the SCGLE theory, whose solution provides the
full time-evolution of the dynamic correlation functions $F_{lm}(k;\tau)$
and $F^S_{lm}(k;\tau)$ and of the memory functions
$\Delta\zeta^{*}_{\alpha}(\tau)$. These equations may be
numerically solved using standard methods once the projections
$S_{lm}(k)$ of the static structure factor are provided. Under
some circumstances, however, one may only be interested in
identifying and locating the regions in state space that
correspond to the various possible ergodic or non ergodic phases
involving the translational and orientational degrees of freedom
of a given system. For this purpose it is possible to derive from the
full SCGLE equations the so-called bifurcation equations, i.e.,
the equations for the long-time stationary solutions of equations
(\ref{flmqframe1})-(\ref{zetarqq}). These are written in terms of
the so-called non-ergodicity parameters, defined as

\begin{equation}
f_{lm}(k)\equiv \lim_{\tau \to
\infty}\frac{F_{lm}(k;\tau)}{S_{lm}(k)},\label{fc0}
\end{equation}

\begin{equation}
f^S_{lm}(k)\equiv \lim_{\tau \to
\infty}F^S_{lm}(k;\tau),\label{fs0}
\end{equation}
and
\begin{equation}
\Delta\zeta_{\alpha}^{*(\infty)} \equiv \lim_{\tau \to \infty}
\Delta\zeta_{\alpha}^{*}(\tau), \label{dz0}
\end{equation}
with $\alpha= T, R$. The
simplest manner to determine these asymptotic solutions is to take
the long-time limit of Eqs. (\ref{flmqframe1})-(\ref{zetarqq}),
leading to a system of coupled equations for $f_{lm}(k)$, $f^S_{lm}(k)$, and
$\Delta\zeta_{\alpha}^{*(\infty)}$.

It is not difficult to show that the resulting equations  can be
written as
\begin{equation}
f_{lm}(k)=\frac{\left[S_{lm}(k)\right]
\lambda_T^{(lm)}(k)\lambda_R^{(lm)}(k)}{S_{lm}(k)\lambda_T^{(lm)}(k)
\lambda_R^{(lm)}(k)+k^2\gamma_T\lambda_R^{(lm)}(k)+l(l+1)\gamma_R\lambda_T^{(lm)}(k)}\label{fc1}
\end{equation}
and
\begin{equation}
f^S_{lm}(k)=\frac{\lambda_T^{(lm)}(k)\lambda_R^{(lm)}(k)}
{\lambda_T^{(lm)}(k)\lambda_R^{(lm)}(k)+k^2\gamma_T\lambda_R^{(lm)}(k)+l(l+1)
\gamma_R\lambda_T^{(lm)}(k)}\label{fs1},
\end{equation}
where the dynamic order parameters $\gamma_T$ and $\gamma_R$,
defined as
\begin{equation}
\gamma_{\alpha}
\equiv\frac{D^0_{\alpha}}{\Delta\zeta_{\alpha}^{*(\infty)}},
\label{parameters}
\end{equation}
are determined from the solution of
\begin{equation}
\frac{1}{\gamma_T}=\frac{1}{6\pi^2n}\displaystyle{\int_0^\infty}
dk \, k^4\sum_{l}[2l+1]\left[1-S^{-1}_{l0}(k)\right]^2
S_{l0}(k)f^S_{l0}(k)f_{l0}(k),\label{gat}
\end{equation}
and
\begin{equation}
\frac{1}{\gamma_R}=\frac{1}{16\pi^2n}\displaystyle{\int_0^\infty}
dk
k^2\sum_{lm}[2l+1][S_{l0}(k)-1]^2S^{-1}_{lm}(k)f^S_{lm}(k)f_{lm}(k)A_{l;0m}^2
.\label{gar}
\end{equation}

As discussed in Ref.  \cite{gory1}, fully ergodic states are
described by the condition that the non-ergodicity parameters
(i.e., $f_{lm}(k)$, $f^S_{lm}(k)$, and
$\Delta\zeta_{\alpha}^{*(\infty)}$) are all zero, and hence, the
dynamic order parameters $\gamma_T$ and $\gamma_R$ are both
infinite. Any other possible solution of these bifurcation
equations indicate total or partial loss of ergodicity. Thus,
$\gamma_T$ and $\gamma_R$ finite indicate full dynamic arrest
whereas $\gamma_T$ finite and $\gamma_R=\infty$ corresponds to the
mixed state in which the translational degrees of freedom are
dynamically arrested but not the orientational degrees of freedom.

\section{Non-equilibrium extension} \label{nonequilib}

The main reason for this brief summary of the SCGLE theory for liquids
with non-spherical inter-particle interactions, is that this
equilibrium theory contains the fundamental ingredients to develop
a theoretical description of the genuine  non-equilibrium
non-stationary irreversible processes characteristic of glassy
behavior, such as aging \cite{nescgle1}.
Thus, let us now outline the conceptual basis and the main steps in the
derivation of the non-equilibrium version of the SCGLE theory for
glass-forming liquids of non-spherical particles, which we shall refer to
as the \emph{non-equilibrium} generalized Langevin equation
(NE-SCGLE) theory.

Our starting point is the non-stationary version \cite{nescgle1}
of Onsager's theory of thermal fluctuations and irreversible
processes \cite{onsager1,
onsager2,onsagermachlup1,onsagermachlup2}, which states that:

\noindent $(I)$ the mean value ${\overline {\textbf a }}(t)$ of the vector $
\mathbf{a}(t)=\left[ a_{1}(t), a_{2}(t),..., a_{\nu
}(t)\right]^{\dagger} $  formed by the $\nu$ macroscopic variables
that describe the state of the system is the solution of some
generally nonlinear equation, represented by
\begin{equation}
\frac{d{\overline {\textbf a }}(t)}{dt}=
\mathcal{R}\left[{\overline {\textbf a }}(t)   \right],
\label{releq0}
\end{equation}
whose linear version in the vicinity of a stationary state ${\overline {\textbf a }}^{ss} $
(i.e.,  $\mathcal{R}\left[{\overline {\textbf a }} ^{ss} \right]=0$) reads
\begin{equation}
\frac{d{\Delta\overline {\textbf a }}(t)}{dt}= -\mathcal{L} [{\overline {\textbf a }}^{ss}]
\cdot \mathcal{E}
\left[{\overline {\textbf a }}^{ss}  \right] \cdot \Delta\overline {\textbf a }(t),
\label{releq1}
\end{equation}
with $\Delta\overline {\textbf a }(t)\equiv \overline {\textbf a }(t)-\overline {\textbf a }^{ss}$, and that:

\noindent  $(II)$ the relaxation equation for the $\nu\times \nu$
covariance matrix $\sigma (t) \equiv \overline{ \delta{\textbf a
}(t)\delta{\textbf a }^\dagger(t)}$ of the non-stationary
fluctuations  $\delta {\textbf a }(t) \equiv {\textbf a
}(t)-{\overline {\textbf a }}(t)$  can be written  as
\cite{nescgle1}
\begin{equation} \label{sigmadtirrev2}
\frac{d\sigma(t)}{dt} = -\mathcal{L} [{\overline {\textbf a }}(t)]
\cdot \mathcal{E}\left[{\overline {\textbf a }}(t)\right]\cdot
\sigma (t)
\end{equation}
\begin{equation} \nonumber
- \sigma (t) \cdot \mathcal{E}\left[{\overline{\textbf
a}}(t)\right] \cdot\mathcal{L}^{\dagger } [{\overline{\textbf
a}}(t)] + \left( \mathcal{L} [{\overline{\textbf a}}(t)] +
\mathcal{L}^{\dagger } [{\overline{\textbf a}}(t)] \right).
\end{equation}

In these equations $\mathcal{L} [{\textbf a}]$ is a $\nu \times \nu$
``\emph{kinetic}'' matrix, defined in terms of $\mathcal{R}\left[{\textbf a}\right]$ as $\mathcal{L} [{\textbf a}]\equiv
-\left(
\partial \mathcal{R}\left[
{\textbf a }\right]/ \partial {\textbf a }\right) \cdot
\mathcal{E}^{-1}\left[{\textbf a}\right]$,  whereas $\mathcal{E}\left[{\textbf a }\right]$ is the $\nu
\times \nu$ thermodynamic (``stability'') matrix, defined as
\begin{equation}
\mathcal{E}_{ij}[{\textbf a }] \equiv -\frac{1}{k_B}\left(
\frac{\partial^2 S[{\textbf a }]}{\partial a_i\partial a_j}
\right)= -\left( \frac{\partial F_i[{\textbf a }]}{\partial a_j}
\right) \ \ \ \ (i,j=1,2,...,\nu), \label{matrixE}
\end{equation}
with $S[{\textbf a }]$ being the entropy  and  $ F_j[
{\textbf a}] \equiv k_B^{-1} \left(
\partial S[ {\textbf a}] /
\partial a_j \right)$ the conjugate intensive variable associated with
$a_j$. The function $S=S[{\textbf a }]$, which assigns a value of
the entropy $S$ to any possible state point {\textbf a} in the
state space of the system, is thus the so-called fundamental
thermodynamic relation \cite{callen}, and constitutes the most important and fundamental
external input of the non-equilibrium theory. The previous equations, however, do not
explicitly require the function $S=S[{\textbf a }]$, but only its second derivatives
defining the stability matrix $\mathcal{E}[{\textbf a }]$. The most important property
of the matrix $\mathcal{E}[{\textbf a }]$ is that its inverse is the covariance of the
\emph{equilibrium} fluctuations, i.e.,
\begin{equation}
\mathcal{E}[ \overline{{\textbf a }}^{eq}] \cdot \sigma^{eq}=I \label{oz},
\end{equation}
with $\sigma^{eq}_{ij}\equiv \overline{\delta a_i\ \delta a_j }^{eq}$, where
the average is taken with the probability distribution $P^{eq}[{\textbf a }]$
of the equilibrium ensemble.

In addition, the non-equilibrium version of Onsager's formalism
introduces the globally non-stationary (but locally stationary)
extension \cite{nescgle1} of the generalized Langevin equation for
the stochastic variables $\delta a_{i}(t+\tau) \equiv
a_{i}(t+\tau)- {\overline a_i}(t)$ \cite{nescgle1},
\begin{eqnarray}
\begin{split}\label{legnegle}
\frac{\partial \delta \textbf{a}(t+\tau)}{\partial \tau}=  & -
\omega[\overline{\textbf{a}}(t)]\cdot \sigma^{-1} (t)\cdot
\delta \textbf{a}(t+\tau) \\
&  -\int_0^\tau d\tau '
\gamma[\tau-\tau';\overline{\textbf{a}}(t)] \cdot
\sigma^{-1}(t)\cdot  \delta \textbf{a}(t+\tau') +
\textbf{f}(t+\tau),
\end{split}
\end{eqnarray}
where the random term $\textbf{f}(t+\tau)$ has zero mean and
two-time correlation function given by the fluctuation-dissipation
relation
$<\textbf{f}(t+\tau)\textbf{f}(t+\tau')>=\gamma[\tau-\tau';\overline{\textbf{a}}(t)]$.
From this equation one derives the time-evolution equation for the
non-stationary time-correlation matrix $C(\tau;t)\equiv \overline{
\delta{\textbf a }(t+\tau)\delta{\textbf a }^\dagger(t)}$, reading
\begin{eqnarray}
\begin{split}\label{fluctuations3}
\frac{\partial C(\tau;t)}{\partial \tau}=  & -
\omega[\overline{\textbf{a}}(t)]\cdot \sigma^{-1} (t)\cdot
C(\tau;t) \\
&  -\int_0^\tau d\tau '
\gamma[\tau-\tau';\overline{\textbf{a}}(t)] \cdot
\sigma^{-1}(t)\cdot C(\tau';t),
\end{split}
\end{eqnarray}
whose initial condition is $C(\tau=0;t)=\sigma(t)$.
In these equations, $\omega[ {\textbf a }]$ represents conservative
(mechanical, geometrical, or streaming) relaxation processes, and
is just the antisymmetric part of $\mathcal{L} [ {\textbf a }]$,
i.e., $\omega[ {\textbf a }]=(\mathcal{L} [ {\textbf a
}]-\mathcal{L}^{\dagger} [ {\textbf a }])/2$. The memory function
$\gamma[\tau;\overline{\textbf{a}}(t)]$, on the other hand,
summarizes the effects of all the complex dissipative irreversible
processes taking place in the system.

Taking the Laplace transform (LT) of   Eq. (\ref{fluctuations3})
to integrate out the variable $\tau$ in favor of the variable $z$,
rewrites this equation as
\begin{equation}
C(z;t) = \left\{z\mathbf{I} +\mathbf{L}[z;\overline{\textbf{a}}(t)] \cdot
\sigma^{-1}(t)\right\}^{-1}\cdot C(\tau=0;t) \label{cdzt}
\end{equation}
with $\mathbf{L}[z;\overline{\textbf{a}}(t)]$ being the LT of
\begin{equation}
\mathbf{L}[\tau;\overline{\textbf{a}}(t)] \equiv    2\delta (\tau) \omega[\overline{\textbf{a}}(t)] +
\gamma[\tau;\overline{\textbf{a}}(t)].
\end{equation}
To avoid confusion, let us mention that $\mathbf{L}[z;\overline{\textbf{a}}(t)]$
thus defined is not, of course, an angular momentum. In terms of
$\mathbf{L}[z;\overline{\textbf{a}}(t)]$, the phenomenological
``\emph{kinetic}'' matrix $\mathcal{L} [\overline{\textbf{a}}(t)]$
appearing in Eq. (\ref{sigmadtirrev2}), is given by the following relation
\begin{equation}
\mathcal{L}[\overline{\textbf{a}}(t)]=\mathbf{L}[z=0;\overline{\textbf{a}}(t)] \equiv \omega[\overline{\textbf{a}}(t)]
+ \int_0^{\infty} d\tau \gamma[\tau;\overline{\textbf{a}}(t)], \label{okcssppp}
\end{equation}
which extends to non-equilibrium conditions the well-known Kubo formula.
The exact determination of $\gamma[\tau;\textbf{a}]$ is perhaps
impossible except in specific cases or limits; otherwise one must
resort to approximations. These may have the form of a closure
relation expressing $\gamma[\tau;{\overline {\textbf a }}(t)]$ in
terms of the two-time correlation matrix $C(\tau;t)$ itself,
giving rise to a self-consistent system of equations, as we
illustrate in the application below.

These general and abstract concepts have specific and concrete
manifestations, which we now discuss in the particular context of
the description of non-equilibrium diffusive processes in
colloidal dispersions. For this, let us identify the abstract
state variables $a_i$ with the number concentration
$a_{(r,\bm{\Omega})}\equiv N_{(r,\bm{\Omega})}/ \Delta V$ of
particles with orientation $\bm{\Omega}$ in the $r$th cell of an
imaginary partitioning of the volume occupied by the liquid in $C$
cells of volume $\Delta V$. In the continuum limit, the components
of the state vector $\textbf{a}(t)$ then become the microscopic
local concentration profile $n(\textbf{r},\bm{\Omega};t)$ defined
in Eq. (\ref{localcon}) and the \emph{fundamental thermodynamic
relation} $S=S[\textbf{a}]$ (which assigns a value of the entropy
$S$ to any point \textbf{a} of the thermodynamic state space
\cite{callen}) becomes the functional dependence
$S=S[\textbf{n}]$ of the entropy (or equivalently, of the free
energy) on the local concentration profile
$n(\textbf{r},\bm{\Omega};t)$.

Using this identification in Eqs. (\ref{releq0}) and
(\ref{sigmadtirrev2}) leads to the time evolution equations for
the mean value $\overline{n}(\textbf{r},\bm{\Omega};t)$ and for
the covariance
$\sigma(\textbf{r},\bm{\Omega};\textbf{r}'\bm{\Omega}';t)\equiv
\overline{\delta n (\textbf{r},\bm{\Omega};t)\delta n
(\textbf{r}',\bm{\Omega}';t)}$ of the fluctuations $\delta
n(\textbf{r},\bm{\Omega};t) = n(\textbf{r},\bm{\Omega};t)-
\overline{n}(\textbf{r},\bm{\Omega};t)$ of the local concentration
profile $n(\textbf{r},\bm{\Omega};t)$. These two equations are the
non-spherical extensions of Eqs. (3.6) and (3.8)   of Ref.
\cite{nescgle1}, which are  coupled between them through two (translational and
rotational) local mobility functions, $b^T(\textbf{r},\bm{\Omega};t)$ and
$b^R(\textbf{r},\bm{\Omega};t)$, which in their turn, can be written approximately
in terms of the two-time correlation function $C(\textbf{r},\bm{\Omega};\textbf{r}'\bm{\Omega}';t,t')\equiv
\overline{\delta n (\textbf{r},\bm{\Omega};t)\delta n
(\textbf{r}',\bm{\Omega}';t')}$. A set of well-defined
approximations on the memory function of
$C(\textbf{r},\bm{\Omega};\textbf{r}'\bm{\Omega}';t,t')$, which
extends to non-spherical particles those described in Ref.
\cite{nescgle1} in the context of spherical particles, results in the referred NE-SCGLE theory.

Rather than discussing these general NE-SCGLE equations, let us
now write them explicitly as they apply to a more specific (but
still generic) phenomenon, namely, to a glass-forming liquid of
non-spherical particles subjected to a programmed cooling while
\emph{constrained} to remain spatially \emph{homogeneous} and
\emph{isotropic} with fixed number density $\overline{n}$. Thus,
rather than solving the time-evolution equation for
$\overline{n}({\bf r},\bm{\Omega};t)$, we have that
$\overline{n}({\bf r},\bm{\Omega};t)=\overline{n}$ now becomes a
control parameter. As a result, we only have to solve the
time-evolution equation for the covariance
$\sigma(\textbf{r},\bm{\Omega};\textbf{r}'\bm{\Omega}';t)
=\sigma(\textbf{r}-\textbf{r}',\bm{\Omega},\bm{\Omega}';t)$.
Furthermore, let us only consider the simplest cooling protocol,
namely, the instantaneous temperature quench at $t=0$ from an
arbitrary initial temperature $T_i$ to a final value $T_f$.

At this point let us notice that it is actually more practical to identify
the abstract  vector $\mathbf{a}(t)=\left[ a_{1}(t), a_{2}(t),..., a_{\nu
}(t)\right]^{\dagger} $ of state variables not with the local
concentration $\overline{n}({\bf r},\bm{\Omega};t)$ itself, but
with \emph{only one} of its tensorial modes, so that
$\mathbf{a}(t)=\left[ a_{1}(t)\right] $, with  $a_1\equiv n_{lm}(\mathbf{k},t)$, defined in Eq.
(\ref{nlm}). Under these conditions, the corresponding non-stationary covariance
$\sigma (t)$ is just a scalar, denoted by $S_{lm}(k,t)$, and defined as
\begin{equation}
\sigma (t)=S_{lm}(k,t) \equiv \overline{ \delta n^*_{lm}(\mathbf{k},t)\delta
n_{lm}(\mathbf{k},t)},\label{flm1}
\end{equation}
with $\delta n_{lm}(\mathbf{k},t)\equiv
n_{lm}(\mathbf{k},t)-\overline{ n_{lm}(\mathbf{k},t)}$. In other words,
$S_{lm}(k,t)$ is a diagonal element of the matrix $S_{lm,l'm'}(k,t)\equiv
\overline{ \delta n^*_{lm}(\mathbf{k},t)\delta n_{l'm'}(\mathbf{k},t)}$.
The time-evolution equation of $S_{lm}(k,t)$ then follows from identifying
all the elements of Eq. (\ref{sigmadtirrev2}).

The first of such elements is the thermodynamic matrix
$\mathcal{E}\left[{\textbf a }\right]$, which in this case is also a scalar,
that we shall denote by $\mathcal{E}_{lm}\left[n_{lm}(\mathbf{k})\right]$.
It is defined in terms of the second derivative of the entropy
$S[n_{lm}(\mathbf{k})]$ (in a contracted description in which the only
explicit macroscopic variable is $n_{lm}(\mathbf{k})$) as
\begin{equation}
\mathcal{E}_{lm}\left[n_{lm}(\mathbf{k})\right]\equiv
-\frac{1}{k_B}\left(\frac{d^2S[n_{lm}(\mathbf{k})]}{dn_{lm}^2(\mathbf{k})}\right).
\label{matrixe}
\end{equation}
According to Eq. (\ref{oz}), this thermodynamic property is just the
inverse of the equilibrium value of $S_{lm}^{eq}(k) \equiv 
\overline{ \delta n^*_{lm}(\mathbf{k})\delta
n_{lm}(\mathbf{k})}^{eq}$ of $S_{lm}(k,t)$,
\begin{equation}
\mathcal{E}_{lm}\left[n_{lm}(\mathbf{k})\right]=1/ S_{lm}^{eq}(k). \label{ozlm}
\end{equation}
Let us notice, however, that $\mathcal{E}_{lm}\left[n_{lm}(\mathbf{k})\right]$ is
not just the diagonal element of the matrix $\mathcal{E}_{lm,l'm'}\left[n\right]$,
defined in terms of the second partial derivative of the entropy $S[n]$ (in a
\emph{non}-contracted description in which the explicit macroscopic variables
are \emph{all} the tensorial density modes $n_{lm}(\mathbf{k})$ of the microscopic
one-particle density $n(\mathbf{r},\bm{\Omega};t)$) as
\begin{equation}
\mathcal{E}_{lm,l'm'}\left[n\right]\equiv
-\frac{1}{k_B}\left(\frac{\partial ^2S[n]}{\partial n_{lm}(\mathbf{k})\partial n_{l'm'}(\mathbf{k})}\right).
\end{equation}
However, according again to Eq. (\ref{oz}), the inverse of this matrix yields
the full equilibrium covariance $S_{lm,l'm'}^{eq}(k) \equiv \overline{ \delta
n^*_{lm}(\mathbf{k})\delta n_{l'm'}(\mathbf{k})}^{eq}$, whose diagonal element
$S_{lm}^{eq}(k)$ does determine $\mathcal{E}_{lm}\left[n_{lm}(\mathbf{k})\right]$,
according Eq. (\ref{ozlm}). Let us mention, however, that in reality
$\mathcal{E}_{lm}\left[n_{lm}(\mathbf{k})\right]$ is also a functional of the
spatially non-uniform local temperature field $T(\textbf{r})$. To  indicate
this dependence more explicitly we shall denote the thermodynamic matrix as
$\mathcal{E}_{lm}\left[n_{lm}(\mathbf{k});T\right] $. Here, however, we shall
impose the constraint that at any instant the system is thermally uniform,
$T(\textbf{r})=T$, and instantaneously adjusted to the reservoir temperature $T$,
which will then be a (possibly time-dependent) control parameter $T(t)$.

The second element of Eq. (\ref{sigmadtirrev2}) that we must identify is the
kinetic matrix $\mathcal{L} [{\textbf a }]$. For this, let us first compare
the equilibrium version of Eq. (\ref{cdzt}), namely,
\begin{equation}
C(z) = \left\{z\mathbf{I} +\mathbf{L}[z;\overline{\textbf{a}}] \cdot
\sigma^{-1}\right\}^{-1}\cdot \sigma, \label{cdzteq}
\end{equation}
with its particular case in Eq. (\ref{flmqframe1}), in which the scalars
$F_{lm}(k,z)$ and $S_{lm}(k)$  correspond, respectively, to $C(z)$ and
$\sigma$.  This comparison allows us to identify  $\mathbf{L}[z;\overline{\textbf{a}}]$
with the scalar
\begin{equation}
\left[{\frac{k^2D_T^0}
{1+\Delta\zeta_T^*(z)\lambda^{(lm)}_T(k)}+\frac{l(l+1)D_R^0}
{1+\Delta\zeta_R^*(z)\lambda^{(lm)}_R(k)}}\right]. \label{Ldzeq}
\end{equation}
Extending this identification to non-stationary conditions, we have that
\begin{equation}
\mathbf{L}[z;\overline{\textbf{a}}(t)]=\left[{\frac{k^2D_T^0}
{1+\Delta\zeta_T^*(z;t)\lambda^{(lm)}_T(k;t)}+\frac{l(l+1)D_R^0}
{1+\Delta\zeta_R^*(z;t)\lambda^{(lm)}_R(k;t)}}\right], \label{Ldznoneq}
\end{equation}
where the functions $\lambda^{(lm)}_R(k;t)$ are defined as unity and the
functions  $\lambda^{(lm)}_T(k;t)$ as $\lambda^{(lm)}_T(k;t)=1/[1+( k/k_{c}(t)) ^{2}]$,
where $k_c=1.305 \times k_{max}(t)$, with $k_{max}(t)$ being the position
of the main peak of $S_{00}(k;t)$. The functions $\Delta\zeta_T^*(z;t)$ and
$\Delta\zeta_R^*(z;t)$, to be defined below, are the non-stationary versions
of the functions  $\Delta\zeta_T^*(z)$, and $\Delta\zeta_R^*(z)$.

Since $\mathcal{L}[\overline{\textbf{a}}(t)]=\mathbf{L}[z=0;\overline{\textbf{a}}(t)]$
(see Eq. (\ref{okcssppp})), the general and abstract time-evolution equation in Eq.
(\ref{sigmadtirrev2}) for the non-stationary covariance becomes
\begin{eqnarray}
\frac{\partial S_{lm}(k;t)}{\partial t} = & \displaystyle -2\left[{\frac{k^2D_T^0}
{1+\Delta\zeta_T^*(z=0;t)\lambda^{(lm)}_T(k=0;t)}+\frac{l(l+1)D_R^0}
{1+\Delta\zeta_R^*(z=0;t)\lambda^{(lm)}_R(k=0;t)}}\right] \\& \times
 \left[\mathcal{E}_{lm}(k,t) S_{lm}(k;t) - 1 \right],
\label{relsigmadifnonspher12} \nonumber
\end{eqnarray}
where $\mathcal{E}_{lm}(k,t)=\mathcal{E}_{lm}\left[n_{lm}(\mathbf{k});T(t)\right] $. 
In the present application to the instantaneous isochoric quench at time $t=0$ to a
final temperature $T_f$ and fixed bulk density $n$, this property is a constant, i.e.,  
for $t>0$ we have that $\mathcal{E}_{lm}(k,t)=\mathcal{E}_{lm}\left[n_{lm}(\mathbf{k});T_f\right] 
=\mathcal{E}^{(f)}_{lm}(k)$. In addition, in consistency with the coarse-grained limit 
$z=0$ in $\Delta\zeta_T^*(z=0;t)$ and $\Delta\zeta_R^*(z=0;t)$, we have also approximated
$\lambda^{(lm)}_T(k;t)$ and $\lambda^{(lm)}_R(k;t)$ by its $k\to 0$ limit  
$\lambda^{(lm)}_T(k=0;t)$ and $\lambda^{(lm)}_R(k=0;t)$, which are actually unity. Thus, 
the previous equation reads
\begin{equation}
\frac{\partial S_{lm}(k;t)}{\partial t} = -2\left[ k^2 D^T_0
b^T(t)+l(l+1)D^R_0 b^R(t) \right]
\mathcal{E}_{lm}^{(f)}(k) \left[S_{lm}(k;t) - 1/\mathcal{E}_{lm}^{(f)}(k) \right],
\label{relsigmadifnonspher13}
\end{equation}
where  the translational and rotational time-dependent mobilities $b^T(t)$
and $b^R(t)$ are defined as
\begin{equation}
b^T(t)= [1+\int_0^{\infty} d\tau\Delta{\zeta}^*_T(\tau; t)]^{-1}
\label{btdt}
\end{equation}
and
\begin{equation}
b^R(t)= [1+\int_0^{\infty} d\tau\Delta{\zeta}^*_R(\tau; t)]^{-1}.
\label{brdt}
\end{equation}
in terms of the non-stationary $\tau$-dependent friction functions
$\Delta\zeta^*_T(\tau; t)$ and $\Delta\zeta^*_R(\tau; t)$.

In order to determine  $b^T(t)$ and $b^R(t)$, we adapt to non-equilibrium
non-stationary conditions, the same approximations leading to Eqs. (\ref{zetart})
and (\ref{zetarqq}) for the equilibrium  friction functions
$\Delta\zeta^*_T(\tau)$ and $\Delta\zeta^*_R(\tau)$, which in the present case
lead to similar approximate expressions for $\Delta\zeta^*_T(\tau; t)$ and
$\Delta\zeta^*_R(\tau; t)$, namely,
\begin{equation}
\Delta\zeta^*_T(\tau;t) =\frac{1}{3}\frac{D_T^0}{(2\pi)^3n}\int
d\mathbf{k}
k^2\sum_{l}\left[2l+1\right]\left[1-S^{-1}_{l0}(k;t)\right]^2
F^S_{l0}(k,\tau;t)F_{l0}(k,\tau;t)\label{zetart2}
\end{equation}
and
\begin{equation}
\Delta\zeta^*_R(\tau;t)=\frac{1}{2}\frac{D_R^0}{(2\pi)^3}
\frac{n}{4}\frac{1}{(4\pi)^2}\int
d\mathbf{k}\sum_{lm} \left[2l+1\right]
h^2_{l0}(k;t)\left[A_{l;0m}\right]^2\left[S^{-1}_{lm}(k;t)\right]^2
F^S_{lm}(k,\tau;t)F_{lm}(k,\tau;t),\label{zetarq}
\end{equation}
where $F_{lm;l'm'}(\mathbf{k},\tau; t)$ are the non-stationary,
$\tau$-dependent correlation functions
$F_{lm;l'm'}(\mathbf{k},\tau; t)\equiv\langle \delta
n^*_{lm}(\mathbf{k},t+\tau)\delta n_{l'm'}(\mathbf{k},t)\rangle$,
with $F_{lm;l'm'}^S(\mathbf{k},\tau; t)$ being the corresponding
\emph{self} components.

In a similar manner, the time-evolution equations for
$F_{lm;l'm'}(k,\tau;t)$ and $F_{lm;l'm'}^S(k,\tau; t)$ are
written, in terms of the Laplace transforms $F_{lm;l'm'}(k,z;t)$,
$F_{lm;l'm'}^S(k,z; t)$, $\Delta\zeta_T^*(z;t)$, and
$\Delta\zeta_R^*(z;t)$, as
\begin{equation}
F_{lm}
(k,z;t)=\frac{S_{lm}(k;t)}{z+\displaystyle\frac{k^2D_T^0S^{-1}_{lm}(k;t)}{1+\Delta\zeta_T^*(z;t)
\lambda^{(lm)}_T(k;t)}+\frac{l(l+1)D_R^0S^{-1}_{lm}(k;t)}{1+\Delta\zeta_R^*(z;t)\lambda^{(lm)}_R(k;t)
}},\label{flmqframenonstat}
\end{equation}

\begin{equation}
F_{lm}^S
(k,z;t)=\frac{1}{z+\displaystyle\frac{k^2D_T^0}{1+\Delta\zeta_T^*(z;t)
\lambda^{(lm)}_T(k;t)}+\frac{l(l+1)D_R^0}{1+\Delta\zeta_R^*(z;t)\lambda^{(lm)}_R(k;t)
}}.\label{flmsqframenonstats}
\end{equation}

For given specific thermodynamic functions  
$\mathcal{E}_{lm}\left[n_{lm}(\mathbf{k});T_f\right]$, Eqs.
(\ref{relsigmadifnonspher13})-(\ref{flmsqframenonstats}) constitute
a closed set of equations for the non-equilibrium properties 
$S_{lm}(k;t)$, $F_{lm}(k,\tau;t)$, $F_{lm}^S (k,\tau;t)$, whose
solution provides the NE-SCGLE description of the non-stationary
and non-equilibrium structural relaxation of glass-forming liquids
formed by non-spherical  particles. In a concrete application,
these equations only require as an input the specific form of
$\mathcal{E}_{lm}\left[n_{lm}(\mathbf{k});T_f\right]$
and of the (arbitrary) initial static structure factor projections
$S_{lm}(k)\equiv S_{lm}(k;t=0)$. In the following section we illustrate
the concrete application of the  theory with a simple but interesting application.

\section{Illustrative application: interacting dipoles with random fixed positions.}\label{agingspinglass}

Eqs. (\ref{relsigmadifnonspher13})-(\ref{flmsqframenonstats}) describe
the coupled translational and rotational dynamics of a Brownian liquid
of non-spherical particles in search of thermodynamic equilibrium after
a sudden quench. A thorough application to a concrete system should then
exhibit the full interplay of the translational and
rotational degrees of freedom during this process. As mentioned in the
introduction, however, carrying out such an exercise falls out of the scope of the
present paper. Instead, as an illustrative application here
we discuss the solution of our resulting equations describing
the irreversible evolution of the orientational dynamics of a system of
strongly interacting dipoles with fixed but random positions
subjected to a sudden temperature quench.

For this, let us recall that two important inputs of Eqs.
(\ref{relsigmadifnonspher13})-(\ref{flmsqframenonstats}), are the
short-time self-diffusion coefficients $D^0_T$ and $D^0_R$, which
describe, respectively,  the short-time Brownian motion of the center
of mass and of the orientations of the particles. Hence, \emph{arbitrarly} setting
$D^0_T=0$ implies that the particles are prevented from diffusing
translationally in any time scale, thus remaining fixed
in space.  Within this simplification Eq. (\ref{relsigmadifnonspher13})
reduces to
\begin{equation}
\frac{\partial S_{lm}(k;t)}{\partial t} = -2
l(l+1)D^R_0 b^R(t)\mathcal{E}_{lm}^{(f)}(k)
\left[S_{lm}(k;t) -1/\mathcal{E}_{lm}^{(f)}(k)\right],
\label{relsigmadifnonspher}
\end{equation}
whereas Eqs. (\ref{flmqframenonstat}) and
(\ref{flmsqframenonstats}) now read
\begin{equation}
F_{lm}
(k,z;t)=\frac{S_{lm}(k;t)}{z+\displaystyle\frac{l(l+1)D_R^0S^{-1}_{lm}(k;t)}{1+\Delta\zeta_R^*(z;t)\lambda^{(lm)}_R(k,t)
}},\label{flmqframenonstat2}
\end{equation}
and
\begin{equation}
F_{lm}^S
(k,z;t)=\frac{1}{z+\displaystyle\frac{l(l+1)D_R^0}{1+\Delta\zeta_R^*(z;t)\lambda^{(lm)}_R(k,t)
}}.\label{flmsqframenonstat2}
\end{equation}
Also, the time-dependent translational mobility satisfies
$b^T(t)=1$. Hence, we only need to complement Eqs.
(\ref{relsigmadifnonspher}), (\ref{flmqframenonstat2}) and
(\ref{flmsqframenonstat2}) with
\begin{equation}
b^R(t)= [1+\int_0^{\infty} d\tau\Delta{\zeta}^*_R(\tau; t)]^{-1}.
\label{brdt2}
\end{equation}
and
\begin{equation}
\Delta\zeta^*_R(\tau;t)=\frac{1}{2}\frac{D_R^0}{(2\pi)^3}
\frac{n}{4}\frac{1}{(4\pi)^2}\int d\mathbf{k}\sum_{lm}
\left[2l+1\right]
h^2_{l0}(k;t)\left[A_{l;0m}\right]^2\left[S^{-1}_{lm}(k;t)\right]^2
F^S_{lm}(k,\tau;t)F_{lm}(k,\tau;t),\label{zetarq2}
\end{equation}
where $A_{l;0m}\equiv\left[C_{l0}^+\delta_{1,m}+C_{l0}^-\delta_{-1,m}\right]$
and $C_{l0}^{\pm}\equiv\sqrt{(l \mp 0)(l+1)}$.
In the following subsections we report the simplest application of these equations.

\subsection{The dipolar hard-sphere liquid with frozen positions.}

Let us consider a system formed by \emph{N} identical dipolar hard
spheres of diameter $\sigma$ bearing a point dipole of magnitude $\mu$
in their center, such that the dipolar moment of the $n$-th particle
$(n=1,2 ...N)$ can be written as $\bm{\mu}_n=\mu\hat{\bm{\mu}}_n$
where the unitary vector $\hat{\bm{\mu}}_n$ describes its orientation.
Thus, the orientational degrees of freedom of the system, $\bm{\Omega}^N$,
are described by the set of unitary vectors
$(\hat{\bm{\mu}}_1,\hat{\bm{\mu}}_2,...,\hat{\bm{\mu}}_N)=\bm{\Omega}^N$,
so that the pair potential $u(\mathbf{r}_n,\mathbf{r}_{n'};\bm{\Omega}_n,\bm{\Omega}_{n'})$
between particles $n$ and $n'$ is thus the sum of the radially-symmetric
hard-sphere potential $u_{HS}(|\mathbf{r}_n-\mathbf{r}_{n'}|)$ plus the
dipole-dipole interaction, given by
\begin{eqnarray}
u_{dip}(\mathbf{r}_n,\mathbf{r}_{n'};\bm{\Omega}_n,\bm{\Omega}_{n'})= \mu^2
|\mathbf{r}_n-\mathbf{r}_{n'}|^{-5} [(\mathbf{r}_n-\mathbf{r}_{n'})^2(\hat{\bm{\mu}}_ n\cdot\hat{\bm{\mu}}_{n'})\\\nonumber
-3((\mathbf{r}_n -\mathbf{r}_{n'})\cdot\hat{\bm{\mu}}_n)((\mathbf{r}_n - \mathbf{r}_{n'})\cdot\hat{\bm{\mu}}_{n'})].\label{potiential}
\end{eqnarray}
The state space of this system is spanned by the number density $n$
and the temperature $T$, expressed in dimensionless form as $[n\sigma^3]$
and $[k_BT\sigma^3/\mu^2]$ (with $k_B$ being Boltzmann's constant). From now on 
we shall denote $[n\sigma^3]$ and $[k_BT\sigma^3/\mu^2]$ simply as $n$ and $T$, 
i.e., we shall use $\sigma$ as the unit of length, and $\mu^2/k_B\sigma^3$ as 
the unit of temperature; most frequently, however, we shall also refer to the
hard-sphere volume fraction $\phi\equiv \pi n/6$.

The application of the NE-SCGLE equations starts with the external
determination of the thermodynamic function
$\mathcal{E}_{lm}^{(f)}(k)\equiv\mathcal{E}_{lm}(k;\phi,T_f)$. At a
given state point $(\phi,T)$ the function $\mathcal{E}_{lm}(k;\phi,T)$
can be determined using the fact that its inverse is identical to the
projection $S_{lm}^{eq}(k;\phi,T)$ of the \emph{equilibrium } static
structure factor $S^{eq}(\mathbf{k},\bm{\mu},\bm{\mu}')$ at that state
point. In the context of the present application, this equilibrium property
will be approximated by the solution of the mean spherical approximation (MSA)
for the dipolar hard sphere (DHS) fluid developed by Wertheim \cite{wertheim}.
The details involved in the determination of the resulting equilibrium static
structure factor, whose  only non zero projections are
$S_{00}^{eq}(k), S_{10}^{eq}(k)$ and $S_{11}^{eq}(k)=S_{1-1}^{eq}(k)$, can
be consulted in Ref. \cite{schilling1}.

\begin{figure}
{\includegraphics[width=3.5in, height=3.5in]{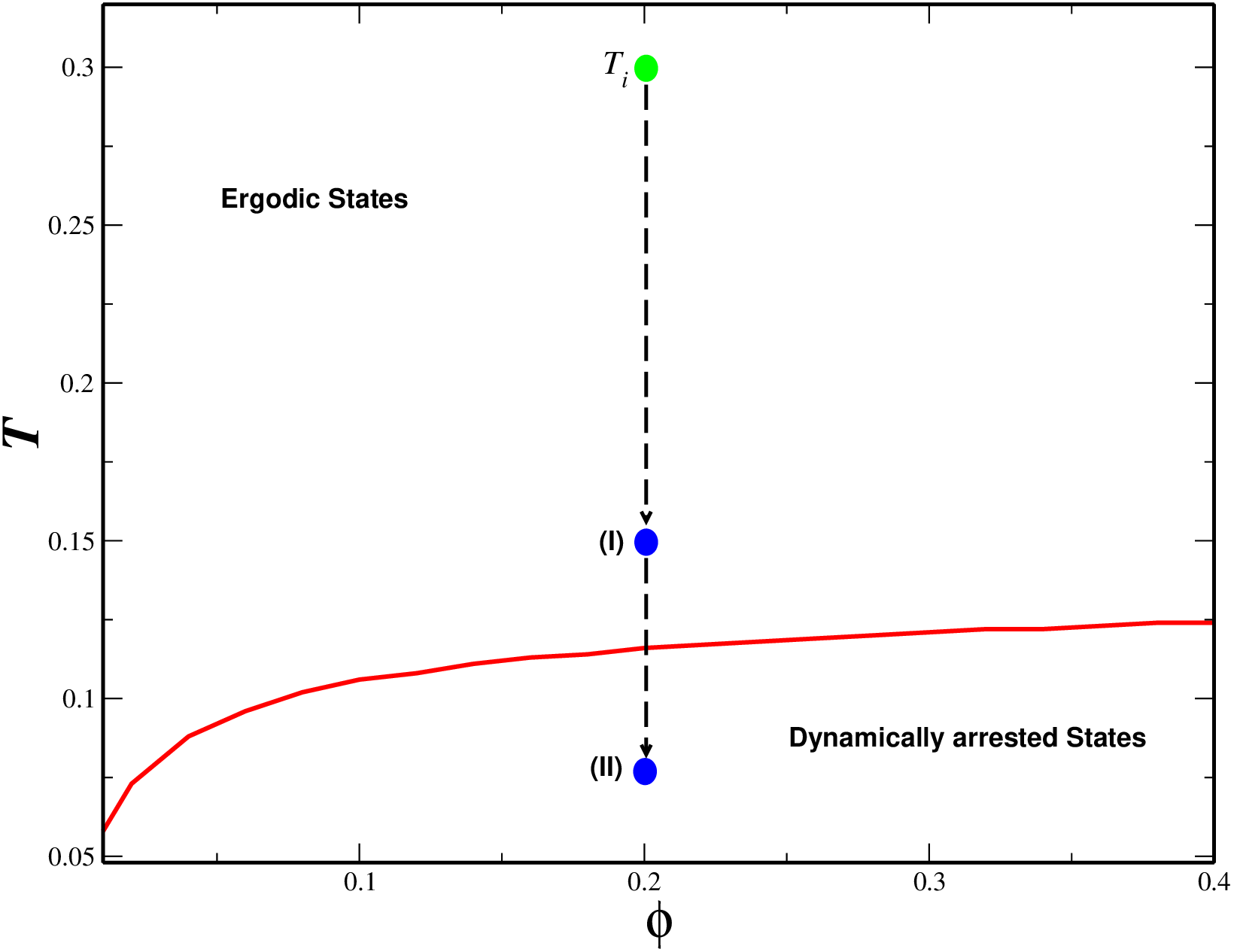}}
\caption{Dynamical arrest line (solid curve) in the ($\phi,T$) state space of
the system of interacting dipoles with fixed positions. This line is the 
boundary between the region of ergodic states, at which the system is predicted
to reach thermodynamic equilibrium, and the predicted region of dynamically 
arrested states. Each of the two superimposed vertical dashed arrows represent
the quench of the system from an initial temperature $T_i$ (green dot) to a final
temperature $T_f$  (blue dots), in one case above ($\mathbf{I}$) and in the other
case below  ($\mathbf{II}$) the dynamic arrest line.} \label{diagram}
\end{figure}

The equilibrium projections $S_{lm}^{eq}(k;\phi,T)$ can also be used in the
so-called bifurcation equations of the equilibrium theory. These are Eqs.
(\ref{fc1})-(\ref{gar}) for the non-ergodicity parameters
$\gamma_T^{eq}(\phi,T)$ and $\gamma_R^{eq}(\phi,T)$. According to
Eq.(\ref{parameters}), however, $D^0_T=0$ implies $\gamma_T^{eq}(\phi,T)=0$,
so that in the present case we must only solve Eq. (\ref{gar}) for
$\gamma_R^{eq}(\phi,T)$. If the solution is infinite we say that the asymptotic
stationary state is ergodic, and hence, that at the point $(\phi,T)$ the system
will be able to reach its thermodynamic equilibrium state. If, on the other hand,
$\gamma_R^{eq}(\phi,T)$ turns out to be finite, the system is predicted to become
dynamically arrested and thus, the long time limit of $S_{lm}(k,t)$ will differ
from the thermodynamic equilibrium value $S_{lm}^{eq}(k;\phi,T)$. The application
of this criterion leads to the prediction that the system under consideration will
equilibrate for temperatures $T$ above a critical value $T_c(\phi)$, whereas the
system will be dynamically arrested for temperatures below $T_c$. In this manner
one can trace the dynamic arrest line $T_c=T_c(\phi)$, which for our illustrative
example is presented in Fig. \ref{diagram}. For example, along the isochore $\phi=0.2$,
this procedure determines that $T_c=T_c(\phi=0.2)=0.116$.

We can now use the same thermodynamic function
$\mathcal{E}_{lm}^{(f)}(k)\equiv\mathcal{E}_{lm}(k;\phi,T_f)$ to go beyond
the determination of the  dynamic arrest line $T_c=T_c(\phi)$ by solving 
the set of NE-SCGLE equations (\ref{relsigmadifnonspher})-(\ref{zetarq2})
to  describe the rotational diffusive relaxation of our system. For this,
let us notice that these equations
happen to have the same mathematical structure as the NE-SCGLE equations that
describe the translational diffusion of \emph{spherical} particles
(see, e.g., Eqs. (2.1)-(2.6) of Ref. \cite{nescgle5}). Although the physical
meaning of these two sets of equations is totally different, their
mathematical similarity allows us to implement the same method of solution
described in Ref. \cite{nescgle3}. Thus, we do not provide further details
of the numerical protocol to solve Eqs. (\ref{relsigmadifnonspher})-(\ref{zetarq2}), 
but go directly to illustrate the resulting scenario.

At this point let us notice that there are two possible classes of stationary
solutions of Eq. (\ref{relsigmadifnonspher}).  The first class corresponds to
the long-time asymptotic condition $\lim_{t\to\infty}S_{lm}(k;t) =1/\mathcal{E}_{lm}^{(f)}(k)$,
in which the system is able to reach the thermodynamic equilibrium condition 
$S_{lm}^{eq}(k) =1/\mathcal{E}_{lm}^{(f)}(k)$. Equilibration is thus a sufficient
condition for the stationarity of $S_{lm}(k,t)$. It is, however, not \emph{a 
necessary} condition. Instead, according to  Eq. (\ref{relsigmadifnonspher}),
another sufficient condition for stationarity  is that $\lim_{t\to\infty}b_R(t)=0$.
This is precisely the hallmark of dynamically-arrested states. In what follows we
discuss the phenomenology predicted by the solution of 
Eqs. (\ref{relsigmadifnonspher})-(\ref{zetarq2}) for each 
of these two mutually exclusive possibilities.

\subsection{Equilibration of the system of interacting dipoles with random fixed positions.}

\begin{figure}
{\includegraphics[width=3in]{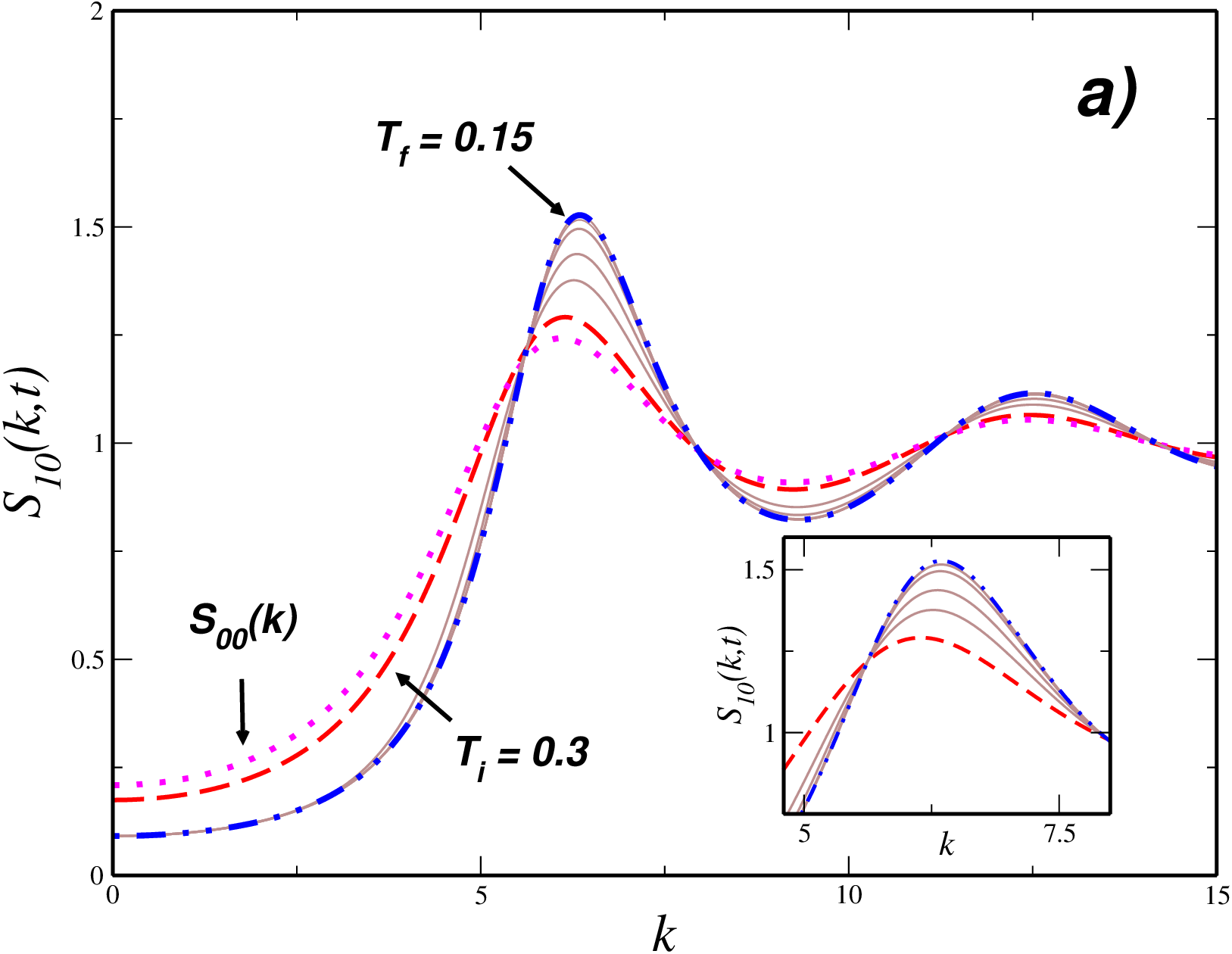}}
{\includegraphics[width=3in]{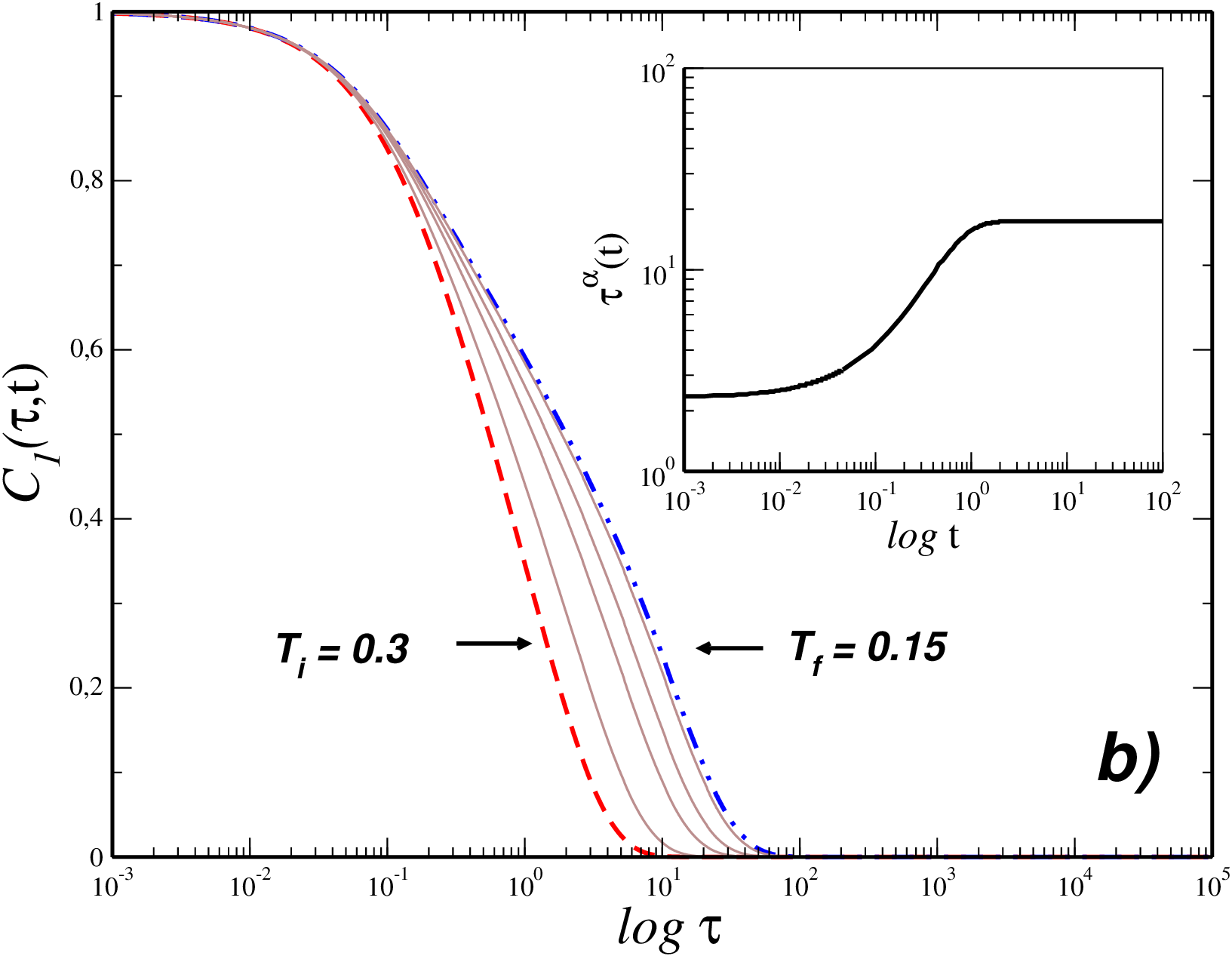}}
\caption{(Color online) Illustration of an equilibration process: (a) Snapshots of the time evolution of the
$l=1,m=0$ static structure factor projection, $S_{10}(k,t)$,
corresponding to the isochoric quench $T_i\to T_f$ for $\phi=0.2$,
with $T_i=0.3$ and $T_f=0.15$. The (red) dashed line is the initial
structure factor $S_{10}(k,t=0)=S_{10}^{(i)}(k)$. The (blue) dot-dashed
line is the asymptotic limit $S_{10}(k,t\to\infty)=S_{10}^{(f)}(k)=S_{10}^{(eq)}(k)$.
The sequence of thinner (brown) solid lines in between represents $S_{10}(k,t)$ for 
$t=0.3,\ 0.78,\ 1.42,\ 2.33\ \text{and}\ t\to\infty$. For reference we also include the non-evolving component $S_{00}(k,t)=S_{00}^{eq}(k;\phi,T_i)$, indicated by the dotted line.  (b) Snapshots of the orientational autocorrelation function $C_{1}(\tau;t)$ as a function of correlation time $\tau$ (thin brown solid lines), corresponding to the same isochoric quench and  same sequence of waiting times $t$ as in (a).  The (red) dashed line represents the initial function $C_{1}(\tau;t=0)=C^{(eq)}_{1}(\tau;\phi,T_i)$ and the (blue) dot-dashed line is the asymptotic limit $C_{1}(\tau;t\to\infty)=C^{(eq)}_{1}(\tau;\phi,T_f)$. The inset plots the $\alpha$-relaxation time, defined as $C_1(\tau_\alpha;t)=1/e$, as a function of  waiting time $t$.}\label{ergo}
\end{figure}

Let us now discuss the  solution of Eqs. (\ref{relsigmadifnonspher})-(\ref{zetarq2})
describing the non-equilibrium response of the system to an instantaneous temperature
quench. For this, we assume that the system was prepared in an equilibrium state 
characterized by the initial value $S_{lm}^{(i)}(k)=S_{lm}^{(eq)}(k,\phi,T_i)=S_{lm}(k,t=0)$, 
of $S_{lm}(k,t)$, and that at time $t=0$ the temperature is instantaneously quenched to a final value $T_f$.
Normally one expects that, as a result, the system will eventually
reach full thermodynamic equilibrium, so that the long time
asymptotic limit of $S_{lm}(k,t)$ will be the equilibrium projections
$S_{lm}^{(eq)}(k;\phi,T_f)$. Such equilibration processes are illustrated in Fig. 
\ref{ergo}(a) with an example in which the system was quenched from an initial equilibrium
state at temperature $T_i=0.3$, $S_{lm}(k,t=0)= S_{lm}(k;\phi,T_i)$, to a final temperature
$T_f=0.15> T_c=0.116$, keeping the volume fraction constant at $\phi=0.2$ (the first of the
two quenches schematically indicated by the dashed vertical arrows of Fig. \ref{diagram}). 

Under these conditions, and from the physical scenario predicted in Fig. 
\ref{diagram},  we should expect that the system will indeed 
equilibrate, so that $S_{lm}(k,t\to \infty)=S_{lm}^{(eq)}(k;\phi,T_f)$. 
This, however, will only be true for $S_{10}(k,t)$ and $S_{11}(k,t)$, 
since, according to Eq. (\ref{relsigmadifnonspher}), $S_{00}(k,t)$ must 
remain constant for $t>0$, indicating that the artificially-quenched 
spatial structure will not evolve as a result of the temperature quench. 
For the same reason, Eqs. (\ref{flmqframenonstat2}) and (\ref{flmsqframenonstat2})
imply that the normalized intermediate scattering functions 
$F_{00}(k,\tau;t)/S_{00}(k;t)$ and $F_{00}^S(k,\tau;t)$ will be unity for 
all positive values of the correlation time $\tau$ and waiting time $t$. 
For reference, the structure of the frozen positions represented by 
$S_{00}(k,t)=S_{00}^{(eq)}(k,\phi,T_i)$,  is displayed in Fig. \ref{ergo}(a) 
by the (magenta) dotted line, which clearly indicates that the fixed positions 
of the dipoles are strongly correlated, in contrast with a system of dipoles 
with purely random fixed positions, in which $S_{00}(k,t)$ would be unity. 
In the same figure, the initial and final equilibrium static structure factor
projections, $S^{(i)}_{10}(k)=S_{10}^{(eq)}(k;\phi, T_i)$ and
$S^{(f)}_{10}(k)=S_{10}^{(eq)}(k;\phi, T_f)$, are represented, respectively,
by the (red) dashed and (blue) dot-dashed curves. The sequence of (brown) 
solid curves in between represents the evolution of $S_{10}(k,t)$ with waiting
time $t$, as a series of snapshots corresponding to the indicated values
of $t$.

For each snapshot of the static structure factor projections $S_{lm}(k,t)$,
the solution of Eqs. (\ref{relsigmadifnonspher})-(\ref{zetarq2}) also
determines a snapshot of each of the dynamic correlation functions $F_{lm}(k,\tau;t)$
and $F^S_{lm}(k,\tau;t)$. These functions are related with other more intuitive and 
experimentally accessible properties, such as the time-dependent autocorrelation function
$C_1(\tau;t)\equiv \langle \sum_{i=1}^N \hat{\bm{\mu}}_i(t+\tau)\cdot\hat{\bm{\mu}}_i(t)\rangle
/\langle \sum_{i=1}^N \hat{\bm{\mu}}_i(t)\cdot\hat{\bm{\mu}}_i(t)\rangle$
of the normalized dipole vectors $\hat{\bm{\mu}}_i$. In fact, since our 
dynamic correlators $F_{lm}(k,\tau;t)$ and $F^{S}_{lm}(k,\tau;t)$ were assumed to be
described from the intermolecular $k$-frame \cite{gory1}, one can relate them
with the time-dependent autocorrelation function $C_1(\tau;t)$ directly through
the following expression \cite{fabbian1},

\begin{equation}
C_{1}(\tau;t)= \frac{1}{3}\lim_{k\to0}\sum_{m=-1}^{1}F^{S}_{1m}(k,\tau;t)\label{ctau}
\end{equation}

Let us notice that, according to Eq. (\ref{flmsqframenonstat2}),
the three terms in the sum on the right hand side of Eq. (\ref{ctau}),
$F^S_{10}(k,\tau;t), F^S_{11}(k,\tau;t)$ and $F^S_{1-1}(k,\tau;t)$,  
satisfy the same equation of motion (which only depends explicitly on $l$)
and thus, contribute exactly in the same manner to the $\tau$ and $t$ 
dependence of $C_{1}(\tau;t)$. Thus, $C_{1}(\tau;t)$ summarizes the 
irreversible time evolution of the orientational dynamics, as 
illustrated in Fig. \ref{ergo}(b) with the snapshots corresponding 
to the same set of evolution times $t$ as the snapshots of $S_{10}(k,t)$ 
in Fig. \ref{ergo}(a).
We observe that $C_{1}(\tau;t)$ starts from its initial equilibrium value,
$C_{1}(\tau;t=0)=C^{(eq)}_{1}(\tau;\phi,T_i)$ and quickly evolves
with waiting time $t$ towards $C_{1}(\tau;t\to\infty)=C^{(eq)}_{1}
(k,\tau;\phi,T_f)$. This indicates that the expected equilibrium
state at ($\phi=0.2,T_f$) is reached without impediment and that the
orientational dynamics remains ergodic at that state point.

As mentioned before, the structure of Eqs.
(\ref{relsigmadifnonspher})-(\ref{zetarq2}) is the same as that of the
equations in \cite{nescgle3} describing the \emph{spherical} case. Thus,
one should not be surprised that the general dynamic and kinetic scenario
predicted in both cases will exhibit quite similar patterns. For example,
the non-equilibrium evolution described by the sequence of snapshots of 
$C_{1}(\tau;t)$ can be summarized by the evolution of its $\alpha$-relaxation
time $\tau_\alpha(t)$, defined through the condition 
$C_{1}(\tau_{\alpha};t)=1/e$. In the inset of Fig. \ref{ergo}(b) we illustrate 
the saturation kinetics of the equilibration process in terms of the $t$-dependence
of  $\tau_\alpha(t)$, as determined from the sequence of snapshots of 
$C_{1}(\tau_{\alpha};t)$ displayed in the figure. Clearly, after a transient
stage, in which  $\tau_\alpha(t)$ evolves from its initial value  
$\tau_\alpha^{eq}(\phi,T_i)$, it eventually saturates to its final
equilibrium value $\tau_\alpha^{eq}(\phi,T_f)$.

\subsection{Aging of the system of interacting dipoles with random fixed positions.}

Let us now present the NE-SCGLE description of the second
class of irreversible isochoric processes, in which the system
starts in an ergodic state but ends in a dynamically arrested state.
For this, let us consider now the case in which the system is subjected
to a sudden isochoric cooling, at fixed volume fraction $\phi=0.2$, and
from the same initial state as before, but this time to the final state 
point $(\phi,T_f=0.095)$ lying inside the region of dynamically arrested 
states (the second of the two quenches schematically indicated by the dashed
vertical arrows of Fig. \ref{diagram}).

\begin{figure}
\includegraphics[width=3in]{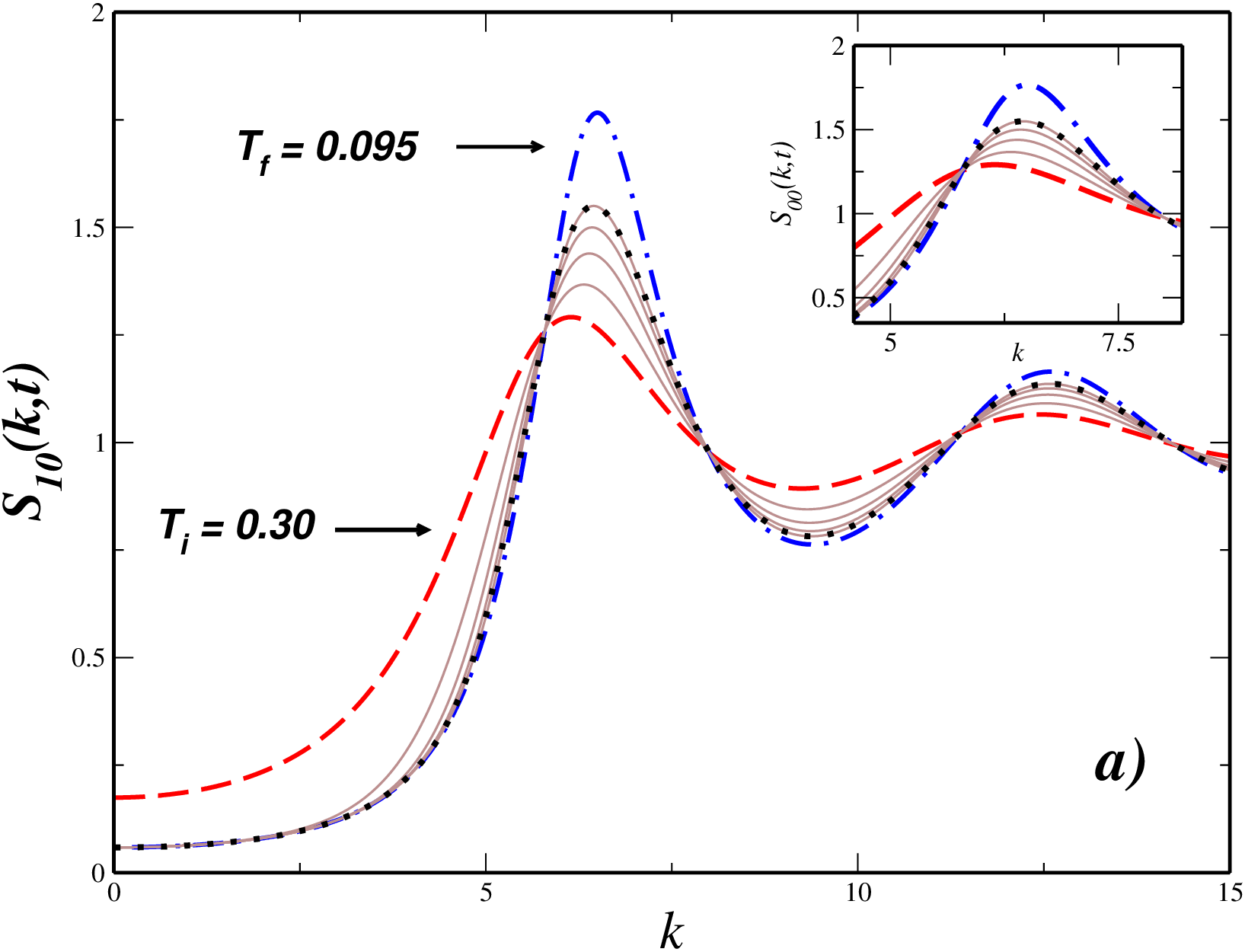}
\includegraphics[width=3in]{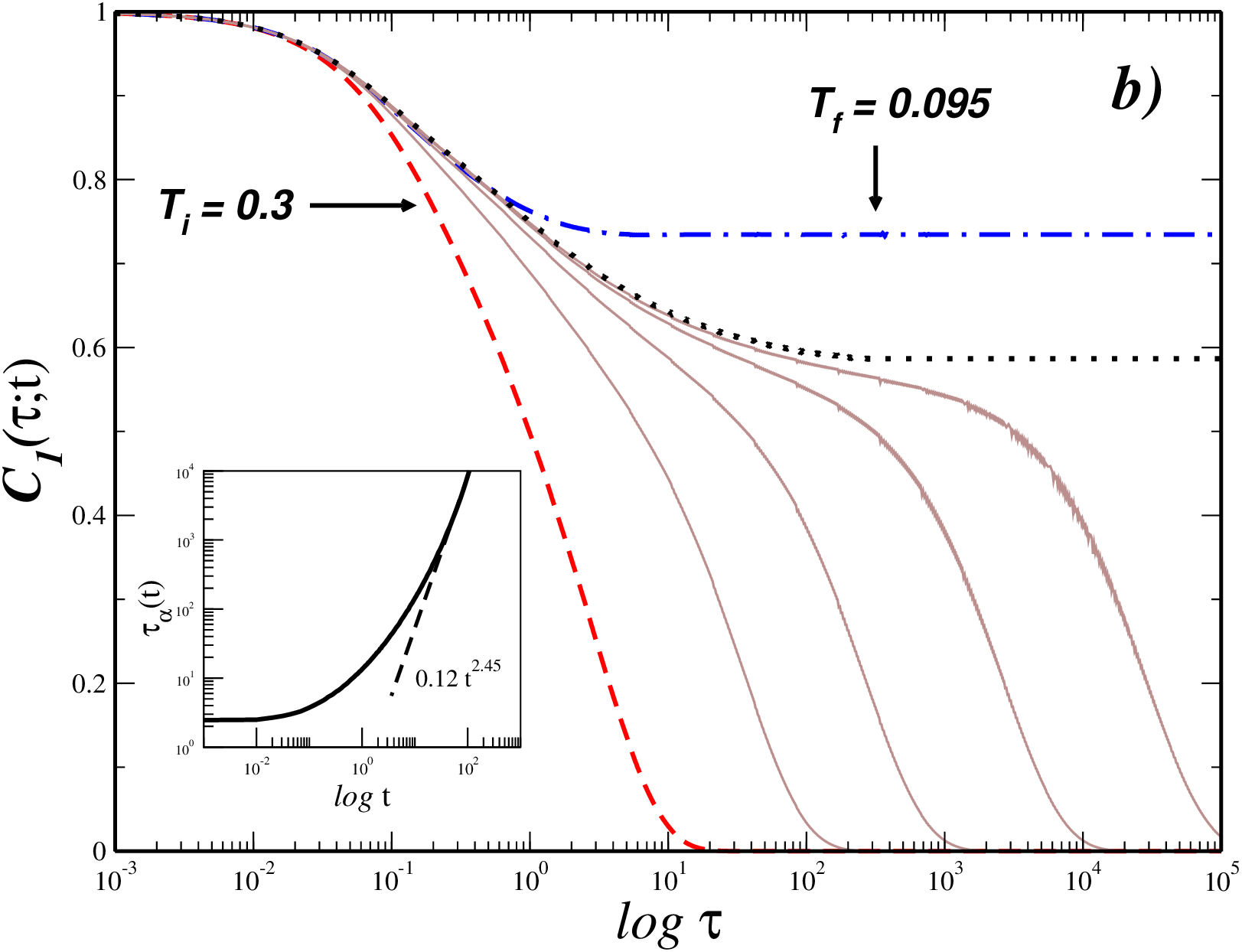}
\caption{Illustration of an aging process: (a) Snapshots of the non-equilibrium time evolution of the
$l=1,m=0$ static structure factor projection, $S_{10}(k,t)$,
corresponding to the isochoric quench $T_i\to T_f$ for $\phi=0.2$,
with $T_i=0.3$ and $T_f=0.095$. The (red) dashed line is the initial
structure factor $S_{10}(k,t=0)=S_{10}^{(i)}(k)$. The (blue) dot-dashed
line is the (now inaccessible) equilibrium structure factor
$S_{10}^{(eq)}(k;\phi,T_f)$,
whereas the (black) dotted line is the predicted asymptotic limit
$S_{10}(k,t\to\infty)=S_{10}^{(a)}(k)$.
The sequence of thinner (brown) solid lines in between represents $S_{10}(k,t)$ for 
$t=1.16,\ 4.264,\ 14.056,\ \text{and}\ 150.61$.  (b) Snapshots of the orientational autocorrelation function $C_{1}(\tau;t)$ as a function of correlation time $\tau$ (thin brown solid lines), corresponding to the same isochoric quench and  same sequence of waiting times $t$ as in (a).  The (red) dashed line represents the initial function $C_{1}(\tau;t=0)=C^{(eq)}_{1}(\tau;\phi,T_i)$, the (blue) dot-dashed
line is the expected (but now inaccessible) equilibrium correlation $C^{(eq)}_{1}(\tau;\phi,T_f)$, and the (black) dotted line
is the predicted long-$t$ asymptotic limit, $C_{1}(\tau;t\to\infty)=C^{(a)}_{1}(\tau)$. The inset plots the corresponding $\alpha$-relaxation
time as a function of  waiting time $t$, with the (black) dashed line representing the asymptotic 
power law $\tau_{\alpha}\propto t^{2.45}$.}\label{sk_arrest}
\end{figure}

Under such conditions, the long-time asymptotic limit of $S_{lm}(k;t)$ will
no longer be the expected equilibrium static structure factor
$S^{(eq)}_{lm}(k;\phi,T_f)$, but another, well-defined non-stationary
structure factor $S^{(a)}_{lm}(k)$. In Fig. \ref{sk_arrest}(a) we illustrate
this behavior with a sequence of snapshots of the non-equilibrium
evolution of $S_{10}(k;t)$ after this isochoric quench at $\phi=0.2$
from $T^{(i)}=0.3$ to $T^{(f)}=0.095$. There we highlight the
initial structure factor
$S_{10}^{(i)}(k)=S_{10}^{(eq)}(k;\phi,T_i)$, represented by the (red)
dashed line and the dynamically arrested long-time asymptotic limit,
$S^{(a)}_{10}(k)$, of the non-equilibrium evolution of $S_{10}(k;t)$,
described by the (black) dotted line. For reference, we also plot the
expected, but inaccessible, equilibrium static structure factor
$S^{(eq)}_{10}(k;\phi,T_f)\ne S^{(a)}_{10}(k)$ (blue dot-dashed line).

Finally, let us illustrate how this scenario of dynamic arrest manifests
itself in the non-equilibrium evolution of the dynamics. We recall that 
for each snapshot of the non-stationary structure factor $S_{lm}(k;t)$, 
the solution of Eqs. (\ref{relsigmadifnonspher})-(\ref{zetarq2}) also 
determines a snapshot of all the dynamic properties at that waiting time 
$t$. For example, in Fig. \ref{sk_arrest}(b) we present the sequence of 
snapshots of $C_{1}(\tau;t)$, plotted as a function of correlation time 
$\tau$, that corresponds to the sequence of snapshots of $S_{10}(k;t)$  
in Fig \ref{sk_arrest}(a). In this figure we highlight in particular the
initial value $C_{1}(\tau;t=0)=C^{(eq)}_{1}(\tau;\phi,T_i)$ (red dashed 
line), the predicted non-equilibrium asymptotic limit, 
$C^{(a)}_{1}(\tau)\equiv \lim_{t\rightarrow \infty} C_{1}(k,\tau;t)$ 
(black dotted line)  and the inaccessible equilibrium value of 
$C^{(eq)}_{1}(\tau;\phi,T_f)$ (blue dot-dashed line). Notice that, 
in contrast with the equilibration process, in which the long-time 
asymptotic solution $C^{(eq)}_{1}(\tau;\phi,T_f)$ decays to zero 
within a finite relaxation time $\tau_\alpha^{eq}(\phi,T_f)$, in 
the present case $C^{(eq)}_{1}(\tau;\phi,T_f)$ does not decay to zero,
but to a finite plateau. This arrested equilibrium correlation function,
however, is completely inaccessible, since now the long-$t$ asymptotic 
limit of $C_{1}(k,\tau;t)$  is $C^{(a)}_{1}(\tau)$, which is also a 
dynamically arrested function, but with a different plateau than 
$C^{(eq)}_{1}(\tau;\phi,T_f)$.

Just like in the equilibration process, which starts at the same 
initial state, here we also observe that at $t=0$, $C_{1}(\tau;t)$ 
shows no trace of dynamic arrest, and that as the waiting time $t$
increases, the relaxation time increases as well. We can summarize
this irreversible evolution of $C_{1}(\tau;t)$ by exhibiting the 
kinetics of the $\alpha$-relaxation time $\tau_\alpha(t)$ extracted 
from the sequence of  snapshots of $C_{1}(\tau_{\alpha};t)$ in the 
same figure. This is done in the inset of fig. \ref{sk_arrest}(b). 
Clearly, after the initial transient stage, in which  $\tau_\alpha(t)$ 
increases from its initial value  $\tau_\alpha^{eq}(\phi,T_i)$ in a 
similar fashion as in the equilibration case, $\tau_\alpha(t)$  no 
longer saturates to any finite stationary value. Instead, it 
increases with $t$ without bound, and actually diverges as a power 
law, $\tau_\alpha(t) \propto t^a$, with $a\approx 2.45$. 

Except for quantitative details, such as the specific value of this
exponent, we find a remarkable general similarity between this predicted
aging scenario of the dynamic arrest of our system of interacting dipoles,
and the corresponding aging scenario of the structural relaxation of a 
soft-sphere glass-forming liquid described in Ref. \cite{nescgle3} (compare,
for example, our Fig. \ref{sk_arrest}(b) above, with Fig. 12 of that reference).
As said above, however, our intention in this paper is not to discuss the 
physics behind these similarities and these scenarios, but only to present
the theoretical machinery that reveals it.

\section{Conclusions}\label{conclusions}

Thus, in summary, we have proposed the extension of the self-consistent
generalized Langevin equation theory for systems of non-spherical
interacting particles (NS-SCGLE), to consider general
non-equilibrium conditions. The main contribution of this work
consist thus in the general theoretical framework, developed in
Sec. \ref{nonequilib}, able to describe the irreversible processes
occurring in a given system after a sudden temperature quench, in which
its spontaneous evolution in search of a thermodynamic equilibrium
state could be interrupted by the appearance of conditions of dynamical
arrest for translational or orientational (or both) degrees of freedom.

Our description consists essentially of the coarse-grained time-evolution
equations for the spherical-harmonics-projections of the static structure factor of the fluid, which involves one translational and one orientational time-dependent mobility functions. These non-equilibrium mobilities, in turn, are determined from the solution of the non-equilibrium version of the SCGLE equations for the non-stationary
dynamic properties (the spherical-harmonics-projections of the self and collective
intermediate scattering functions). The resulting theory is
summarized by Eqs. (\ref{relsigmadifnonspher13})-(\ref{flmsqframenonstats})
which describe the irreversible processes in model liquids of
non-spherical particles, within the constraint that the system remains, on
the average, spatially uniform. This theoretical framework is now ready
to be applied for the description of such nonequilibrium phenomena in many
specific model systems. 

Although in this paper we do not include a thorough discussion of any particular application, in section \ref{agingspinglass} we illustrated the
predictive capability of our resulting equations by applying them
to the description of the isochoric and uniform evolution of
non-equilibrium process of a simple model, namely, a dipolar
hard sphere liquid with fixed random positions,
after being subjected to instantaneous temperature quench. Here we used this example mostly to illustrate some methodological
aspects of the application of the theory, since this specific application allows us to easily implement the numerical methods described in detail in Ref. \cite{nescgle3}.  The same illustrative example, however, also allows us to investigate the relevant features of the orientational dynamics during the equilibration and aging processes, but leaves open many relevant issues, such as the relationship between these predictions and the phenomenology of aging in spin-glass systems. Similarly, the non-equilibrium manifestations of the coupling between translational and rotational dynamics, involved in the complete solution of Eqs. (\ref{relsigmadifnonspher13})-(\ref{flmsqframenonstats}), will be the subject of future communications. Thus, we expect that the general results derived in this paper will be the basis of a rich program of research dealing with these problems.

\section*{Acknowledgments}
This work was supported by the Consejo Nacional de
Ciencia y Tecnolog\'{\i}a (CONACYT, M\'{e}xico) through
grants  No. 242364, No. 182132, No.  237425  and No. 358254. 
L.F.E.A. also acknowledge financial support from Secretar\'ia 
de Educaci\'on P\'ublica (SEP, M\'exico) through PRODEP and 
funding from the German Academic Exchange Service (DAAD)
through the DLR-DAAD programme under grant No. 212.

\end{document}